\documentclass[pra,aps,superscriptaddress,showpacs,floatfix,tightenlines,twocolumn]{revtex4-2}  
\usepackage[T1]{fontenc}
\usepackage[utf8]{inputenc}  
\usepackage[english]{babel}    
\usepackage{graphicx} 
\usepackage{dcolumn}
\usepackage{bm}
\usepackage{float}
\usepackage{amssymb}
\usepackage{amsmath}
\usepackage{url}
\usepackage{algorithm}
\usepackage{algorithmic}
\usepackage{mathbbol}
\usepackage{enumerate}
\usepackage{lipsum}
\usepackage{mathrsfs}
\usepackage{blindtext}
\usepackage{appendix}
\usepackage[hyperindex,breaklinks]{hyperref}
\usepackage{enumitem}
\usepackage{makecell}
\usepackage{booktabs}
\usepackage{multirow}
\usepackage{tabularx}
\usepackage{array}
\usepackage{subfigure}
\usepackage{comment}
\usepackage{siunitx}
\usepackage{soul}
\usepackage{makecell}
\allowdisplaybreaks[4]
\newcommand{\ket}[1]{|#1\rangle}
\newcommand{\bra}[1]{\langle #1 |}

\allowdisplaybreaks[3]

\usepackage{xcolor}

\begin{document}

	\title{TRAM: A Transverse Relaxation Time–Aware Qubit Mapping Algorithm for NISQ Devices}

	\author{Yifei Huang}
	\affiliation{College of Mathematics, Taiyuan University of Technology, Taiyuan, 030024, China}

    \author{Pascal Jahan Elahi}
	\affiliation{Pawsey Supercomputing Research Centre, WA 6151, Australia}
    \affiliation{Department of Physics, The University of Western Australia, Perth, WA 6009, Australia}

    \author{Ugo Varetto}
	\affiliation{Pawsey Supercomputing Research Centre, WA 6151, Australia}
    \affiliation{Department of Physics, The University of Western Australia, Perth, WA 6009, Australia}

	\author{Kan He}
	\email{hekan@tyut.edu.cn}
	\affiliation{College of Mathematics, Taiyuan University of Technology, Taiyuan, 030024, China}

    \author{Jinchuan Hou}
	\email{jinchuanhou@aliyun.com}
	\affiliation{College of Mathematics, Taiyuan University of Technology, Taiyuan, 030024, China}

    \author{Shusen Liu}
	\email{shusen.liu@csiro.au}
	\affiliation{Pawsey Supercomputing Research Centre, WA 6151, Australia}
    \affiliation{Department of Physics, The University of Western Australia, Perth, WA 6009, Australia}

	\date{\today }
\begin{abstract}
Noisy intermediate-scale quantum (NISQ) devices impose dual challenges on quantum circuit execution: limited qubit connectivity requires extensive SWAP-gate routing, while time-dependent decoherence progressively degrades quantum information. Existing qubit mapping algorithms optimize for hardware topology and static calibration metrics but systematically neglect transverse relaxation dynamics ($T_2$), creating a fundamental gap between compiler decisions and evolving noise characteristics. We present TRAM (Transverse Relaxation Time-Aware Qubit Mapping), a coherence-guided compilation framework that elevates decoherence mitigation to a primary optimization objective. TRAM integrates calibration-informed community detection to construct noise-resilient qubit partitions, generates time-weighted initial mappings that anticipate coherence decay, and dynamically schedules SWAP operations to minimize cumulative error accumulation. Evaluated on Qiskit-based simulators with realistic noise models, TRAM outperforms SABRE by 3.59\% in fidelity, reduces gate count by 11.49\%, and shortens circuit depth by 12.28\%, establishing coherence-aware optimization as essential for practical quantum compilation in the NISQ era.
\end{abstract}

	\maketitle

	\section{Introduction}
	
	Quantum computing has attracted considerable attention for its potential to deliver exponential speedups in solving certain classically intractable problems. In recent years, IBM released the Condor superconducting quantum processor \cite{condor}, one of the first to reach over 1000 qubits. Google’s Willow superconducting quantum chip \cite{willow} demonstrated stable operations approaching the quantum error correction (QEC) threshold, typically 1\% for surface codes, marking a significant step toward practical fault-tolerant quantum computing.  Meanwhile, the Zuchongzhi-3 superconducting quantum computer  \cite{zuchongzhi}, developed by the University of Science and Technology of China, achieved new performance benchmarks in superconducting quantum computing systems, with its single-qubit gate fidelity reaching 99.90\%, two-qubit gate fidelity reaching 99.62\% and readout fidelity reaching 99.18\%.  
    
    Despite these advancements, current superconducting quantum chips remain classified as noisy intermediate-scale quantum (NISQ) devices \cite{tuixianggan}, which are constrained by low qubit gate fidelity, a relatively limited number of qubits, and high susceptibility to environmental noises. These noises induce decoherence effects such as depolarization, which causes partial loss of quantum information and thereby severely degrades computational accuracy \cite{jingquelv,kandala1}. Although QEC theoretically offers fault tolerance, its immense resource overhead requiring thousands of physical qubits per logical qubit renders it impractical for NISQ devices \cite{jiucuo}.
    
    Current quantum algorithms and NISQ hardware suffer from a fundamental mismatch. Many quantum algorithms are designed based on idealized assumptions: qubits are noise-free, and two-qubit gates can be applied between any pair of logical qubits \cite{tuixianggan}. In contrast, many NISQ devices, particularly those based on superconducting circuits with fixed planar architectures, restrict connectivity by enforcing two-qubit gates exclusively between physically coupled qubits \cite{nash2020quantum}. To address this mismatch, additional routing operations are required to mediate interactions between distant qubits. Among these, the SWAP gate is the canonical example (Fig. \ref{fig-4}), as it exchanges qubit states between neighboring physical qubits, effectively reconfiguring the logical-to-physical mapping to comply with the hardware topology. This routing process, generally referred to as the qubit mapping (qubit routing) problem, has been proven to be NP-complete \cite{npc}.
    
    To address this problem, existing research can be broadly classified into two types: (1) Exact mathematical optimization methods: These methods aim to find theoretically optimal solutions through mathematical programming. However, they are usually highly sensitive to circuit scale and solution complexity \cite{ilp,menxishou,rout,time}. (2) Heuristic algorithms methods: These methods obtain approximately optimal solutions by designing targeted search strategies. Their key features are high computational efficiency, but they lack guarantees of global optimality \cite{jiang,luo,zhou,li,codar,qucloud,qucloud+,amer,sabre}.

   Numerous studies have focused on the qubit mapping problem from different perspectives. TSA \cite{jiang} integrates subgraph isomorphism, tabu search, and look-ahead strategies to effectively reduce SWAP gate counts. EffectiveQM \cite{luo} adopts a dual-mode search and potential-oriented scoring mechanism, achieving a substantial reduction in SWAP gate counts. DHA \cite{zhou} applies simulated annealing and adjustment of SWAP gate look-ahead depth to reduce circuit depth. ADAC \cite{li} combines subgraph isomorphism with heuristic routing, achieving nearly 50\% performance improvement on IBM Tokyo. CODAR \cite{codar} is aware of gate duration difference and program context, enabling it to extract more parallelism from programs and speed up the quantum programs. QuCloud and QuCloud$+$ \cite{qucloud, qucloud+} reduce SWAP gates and lower circuit error rates through topology-aware qubit partitioning and mapping. Reinforcement learning method \cite{amer} models mapping as a sequential decision problem to minimize circuit depth. SABRE \cite{sabre} remains a widely adopted baseline for minimizing SWAP gate overhead and circuit depth through topology-aware and heuristic look-ahead strategies. LightSABRE \cite{zou2024lightsabre} significantly improves runtime efficiency and reduces the number of SWAP gates, while supporting quantum compilation optimization for disjoint connectivity graphs and circuits with classical control flow.
   
   In general, existing qubit mapping techniques predominantly rely on hardware topology and static calibration data to guide optimization. Nevertheless, these approaches largely overlook the dynamic deterioration of qubit performance during circuit execution, which prevents them from accurately capturing the cumulative impact of decoherence over time, thereby limiting their applicability in realistic, time-dependent quantum computing environments. Incorporating the transverse relaxation time ($T_2$) as a metric of qubit coherence can optimize qubit selection and task scheduling during the mapping process, thus mitigating the impact of cumulative decoherence on circuit fidelity. The $T_2$ is a characteristic time scale that describes how qubits lose phase coherence and undergo dephasing-induced decoherence due to coupling with their environment. When a qubit interacts with its environment, the phase coherence of its qubit state undergoes irreversible loss, progressively suppressing quantum coherence and driving the system toward a classical probabilistic mixed state \cite{schlosshauer}. This physical mechanism leads to coherence loss in entangled and quantum superposition states, fundamentally limiting the fidelity of quantum circuits \cite{jizhun}. Qubits with longer $T_2$ exhibit superior phase coherence, enabling them to undergo more gates within a given time while maintaining coherence, thereby enhancing the overall fidelity and fault-tolerance of quantum circuits.

The proposed Transverse Relaxation Time-Aware Qubit Mapping (TRAM) algorithm redefines qubit mapping as a noise-aware and time-coupled optimization problem, bridging the gap between hardware calibration data and compiler-level decision-making. Rather than treating connectivity as a static geometric constraint, TRAM models qubit mapping as a dynamic process constrained jointly by \emph{hardware topology, decoherence behaviour, and temporal gate evolution}. This perspective allows the compiler to adapt the logical-to-physical mapping throughout execution while maintaining coherence and minimizing cumulative errors.

\medskip
\noindent\textbf{Core idea.}  
Conventional mapping algorithms such as SABRE primarily minimize routing cost measured by SWAP counts or circuit depth, assuming that all qubits are equally reliable. TRAM departs from this assumption by explicitly incorporating decoherence characteristics, represented by the transverse relaxation time $T_2$, as a central optimization signal. Through calibration-informed partitioning and time-dependent mapping strategies, TRAM aims to preserve coherence where it is most valuable-on qubits and at moments most exposed to noise. The central methodological principle of TRAM is therefore to balance \textit{spatial connectivity} with \textit{temporal coherence preservation}.

\medskip
\noindent\textbf{Methodological overview.}  
TRAM operates as a three-stage framework that hierarchically connects structural, temporal, and adaptive optimization layers:

1. Community Detection-Assisted Quantum Transverse Relaxation Partitioning (CQTP). 
   CQTP constructs a noise-aware abstraction of the hardware by analysing calibration data, including two-qubit gate errors, readout errors, and $T_2$ coherence times, to identify regions of qubits that are both densely connected and homogeneously coherent. The resulting partitions form stable physical subgraphs that serve as the substrate for logical circuit placement. This step transforms raw calibration snapshots into topological zones that maximize connectivity while minimizing cross-boundary decoherence.

2. Time-Weighted Heatmap-Based Initial Mapping (THIM). 
   Based on the partitions produced by CQTP, THIM builds a global view of the temporal structure of the circuit. Each two-qubit interaction is weighted by its position in time: later interactions receive exponentially higher weights, reflecting their greater exposure to decoherence. Using this dynamic weighting, THIM assigns logical qubits to physical qubits to minimize the expected time-integrated noise cost. The mapping therefore anticipates future decoherence rather than reacting to it after routing overheads accumulate.

3. Time-Adaptive Dynamic SWAP (T-SWAP). 
   As the circuit executes, T-SWAP dynamically manages routing by evaluating SWAP insertion points through a heuristic cost function that accounts for both physical proximity and cumulative decoherence. Qubits with stronger noise resilience are preferentially used as communication channels, while overused or fragile qubits are progressively de-prioritized through a mild decay mechanism. This adaptive scheduling suppresses redundant SWAP operations and avoids congestion on error-prone paths, reducing both time and error accumulation.

\medskip
\noindent\textbf{Conceptual contribution.}  
Together, these three components implement a coherent methodological framework in which coherence preservation is treated as a first-class optimization objective. TRAM transforms hardware calibration data into actionable compiler signals: CQTP defines where reliable computation can occur, THIM determines how to map logical information in anticipation of noise exposure, and T-SWAP maintains this alignment adaptively during execution. The overall process continuously balances the trade-off between routing efficiency and coherence longevity, providing a scalable path toward high-fidelity compilation on noisy intermediate-scale quantum hardware.

\medskip
\noindent\textbf{Empirical validation.}  
To evaluate TRAM, we developed a noise-simulated environment using the Qiskit framework \cite{kanazawa2023qiskit}, incorporating depolarizing, dephasing, and amplitude-damping noise models parameterized by real hardware calibration data. Across standard benchmarks, TRAM achieves consistent improvements over SABRE: reducing the number of two-qubit gates by an average of 11.49\%, shortening circuit depth by 12.28\%, and improving overall fidelity by 3.59\%. These results demonstrate that coherence-guided mapping, when informed by real device characteristics, can yield systematic gains without introducing significant computational overhead.

In essence, TRAM embodies a hardware-aware and temporally adaptive philosophy of quantum compilation, one that integrates structural modularity, temporal exposure, and dynamic routing into a unified noise-conscious optimization pipeline. It provides a general framework that can extend beyond current superconducting devices, offering a foundation for mapping strategies in future architectures where coherence remains a constrained and valuable computational resource.

The paper is structured as follows.
Section \ref{sec-pre} introduces the background knowledge used in this paper. Section \ref{sec_XN} presents the motivation. Section \ref{sec-TRAM} elaborates on our proposed algorithm TRAM, including its three sub-algorithms: CQTP in \ref{sec-CQTP}, THIM in \ref{sec-THIM}, and T-SWAP in \ref{sec-TSWAP}. Section \ref{sec-EVAL} evaluates TRAM. Finally, Section \ref{sec-conclusion} summarizes the entire paper and outlines future research directions.

	\section{Background}\label{sec-pre}

    In this section, we will give a brief introduction to quantum mechanics, quantum circuit, quantum program, quantum noise models, and superconducting quantum computer.

       \subsection{Quantum mechanics}

    In quantum mechanics, the state of a system is described by a density matrix $\rho$ acting on a Hilbert space $\mathcal{H}$, satisfying three fundamental properties \cite{nielsen}: (1) Hermiticity: $\rho = \rho^\dagger$, where $\rho^\dagger$ is the conjugate transpose of $\rho$. (2) Normalization: $\text{Tr}(\rho) = 1$. (3) Positivity: $\rho \geq 0$.
    
    Quantum states can be classified as pure or mixed. A pure state can be described by a single ket vector $|\psi\rangle$, with $\langle\psi|\psi\rangle = 1$, and corresponds to a rank-one projection operator $\rho = |\psi\rangle\langle\psi|$, satisfying $\rho^2 = \rho$ or $\rm Tr(\rho^2) = 1$. A mixed state represents partial knowledge of the system, expressed as a convex combination $\rho = \sum_{i} p_i |\phi_i\rangle\langle \phi_i|$ with $\text{Tr}(\rho) = 1$ and $\text{Tr}(\rho^2) < 1$. 

    Superposition describes a state as a linear combination of basis states. A single-qubit pure state can be written as $|\psi\rangle = \alpha|0\rangle + \beta|1\rangle$ with $|\alpha|^2 + |\beta|^2 = 1$. An $n$-qubit system resides in a $2^n$-dimensional Hilbert space, enabling superpositions of $2^n$ basis states and quantum parallelism. Entanglement is a uniquely quantum correlation in multi-qubit systems. An entangled state cannot be factorized into a tensor product of single-qubit states. For example, Bell states such as $|\Phi^+\rangle = (|00\rangle + |11\rangle)/\sqrt{2}$ are entangled \cite{horodecki}. Entanglement can be generated via qubit gates, e.g., applying a Hadamard gate to the first qubit of $|00\rangle$ followed by a CNOT yields $|\Phi^+\rangle$ \cite{tuixianggan}. Entangled states exhibit nonlocal correlations that violate Bell inequalities, demonstrating that quantum correlations cannot be explained by classical local realism.
    
    Measurement is described by a set of operators $\{M_m\}$, where outcome $m$ occurs with probability 
\begin{equation*}
  \begin{aligned}
   p(m) = \bra{\psi} M_m^\dagger M_m \ket{\psi},
  \end{aligned}
\end{equation*}
and the post-measurement state becomes 
\begin{equation*}
  \begin{aligned}
\frac{M_m \ket{\psi}}{\sqrt{\bra{\psi} M_m^\dagger M_m \ket{\psi}}}.
  \end{aligned}
\end{equation*}
Completeness $\sum_m M_m^\dagger M_m = I$ ensures normalized probabilities $\sum_m p(m) = \sum_m \bra{\psi} M_m^\dagger M_m \ket{\psi}=1$.

The fidelity of quantum states characterizes the similarity between quantum states after transmission, operations, or noise interference. For quantum states $\rho$ and $\sigma$, the Uhlmann fidelity is $F(\rho, \sigma) = \rm Tr\left( \sqrt{\sqrt{\rho} \sigma \sqrt{\rho}} \right)$ , where $0 \leq F(\rho,\sigma) \leq 1$. For pure states $|\psi\rangle$ and $|\phi\rangle$, we use the simplified version  $F(|\psi\rangle, |\phi\rangle) = |\langle\psi|\phi\rangle|$ \cite{wilde}. 

The Pauli operators are a set of three Hermitian and unitary matrices, denoted as $X$, $Y$, and $Z$, which form a basis for the space of single-qubit operators. They correspond to quantum state transformations associated with spin measurements along the orthogonal axes of the Bloch sphere. Their matrix representations are given by
\begin{equation*}
  \begin{aligned}
    X = \begin{pmatrix} 0 & 1 \\ 1 & 0 \end{pmatrix}, \quad 
    Y = \begin{pmatrix} 0 & -i \\ i & 0 \end{pmatrix}, \quad 
    Z = \begin{pmatrix} 1 & 0 \\ 0 & -1 \end{pmatrix},
  \end{aligned}
\end{equation*}
where $i$ denotes the imaginary unit ($i^2=-1$). The Pauli operators constitute a complete orthogonal basis for describing single-qubit dynamics, noise processes, and quantum error-correction formulations.

    \subsection{Quantum state tomography}

    In quantum state tomography, the goal is to reconstruct an unknown quantum state $\rho$ from a set of measurement outcomes \cite{13ct-8vtv,qtw}. The Maximum Likelihood Estimation (MLE) approach \cite{2001Measurement} treats the measurement statistics as samples drawn from a multinomial distribution and seeks the physical density matrix $\hat{\rho}_\text{mle}$ (positive semidefinite with unit trace) that maximizes the likelihood of the observed data.  
    Formally, if $\{M_k^{(i)}\}$ denotes the measurement operators for the $i$-th measurement setting and $n_k^{(i)}$ the corresponding outcome counts, the MLE problem can be written as
    \begin{equation}
    \label{eq:mle}
    \hat{\rho}_\text{mle} =
    \underset{\rho \ge 0,\, \mathrm{Tr}\rho = 1}{\operatorname{arg\,max}}
    \prod_{i,k}
    \Big[
    \mathrm{Tr}\!\big( M_k^{(i)} \rho (M_k^{(i)})^\dagger \big)
    \Big]^{n_k^{(i)}},
    \end{equation}
    where $\mathrm{Tr}\!\left( M_k^{(i)} \rho (M_k^{(i)})^\dagger \right)$ gives the predicted probability of outcome $k$ in setting $i$. The optimization therefore identifies the physical density matrix that best explains the observed measurement frequencies within statistical uncertainty.

\subsection{Quantum noise models in quantum computing}

In quantum computing, noise processes are commonly modeled using a set of canonical quantum channels. Among them, depolarizing, dephasing, and amplitude damping channels capture the dominant decoherence mechanisms across most qubit platforms: random state mixing, phase coherence loss, and energy relaxation, respectively. These channels admit a compact representation in terms of Kraus operators, which provides a mathematically consistent description of non-unitary dynamics and facilitates both analytical treatment and efficient simulation.

Depolarizing errors randomly corrupt both amplitude and phase information of a quantum state, without preserving any particular basis, degrading a pure state into a mixed state  \cite{nielsen,tuixianggan}. For a single-qubit system, assuming the depolarization probability is $P$, the depolarizing noise channel is represented by $\{K_i\}_{i=0}^3$, where $K_0=\sqrt{1-P}\mathbb{I}$, $K_1=\sqrt{{P}/{3}}X$, $K_2=\sqrt{{P}/{3}}Y$, $K_3=\sqrt{{P}/{3}}Z$. $\mathbb{I}$ is the identity operator, and $X,Y,Z$ denote Pauli operators. 

Dephasing noise undermines the phase coherence of the quantum state, resulting in the loss of phase information \cite{breuer,hayashi}. For a single-qubit system, assuming the dephasing probability is $\gamma$, the dephasing noise channel is represented by a set of Kraus operators $\{E_i\}_{i=0}^1$, where $E_0=\sqrt{1-\gamma}\mathbb{I}$, $E_1=\sqrt{\gamma}Z$. 

Amplitude damping represents the energy relaxation of a single-qubit quantum state, describing the decay from the state $|1\rangle$ to the state $|0\rangle$  \cite{nielsen,breuer}. For a single-qubit system, assuming the relaxation probability is $\lambda$, the amplitude damping channel is represented by a set of Kraus operators $\{F_i\}_{i=0}^1$, where $F_0=|0\rangle\langle0|+\sqrt{1-\lambda}|1\rangle\langle1|$, $F_1=\sqrt{\lambda}|0\rangle\langle1|$.

    \subsection{Quantum program and quantum circuit}

\begin{figure}[ht]
  \centering
  \subfigure[The quantum circuit of the decomposed Fredkin gate.]{
   \includegraphics[width=245pt]{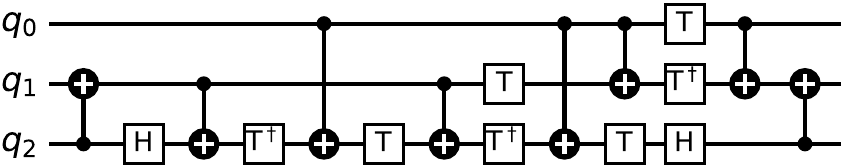}\label{fig-1(a)}
  }
  \hspace{0pt}
  \subfigure[The DAG of the decomposed Fredkin gate.]{
   \includegraphics[width=245pt]{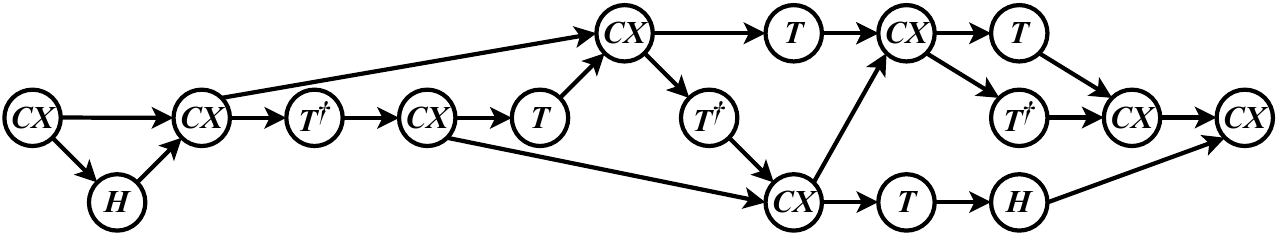}\label{fig-1(b)}
  }
  \caption{\label{fig-1} A quantum circuit and its corresponding DAG.}
\end{figure}

    The quantum program is code written in programming languages to describe quantum computing tasks, incorporating quantum operations and classical control logic. For example, in the Qiskit framework, a quantum circuit is constructed using the QuantumCircuit class \cite{qiskit}. In the following circuit representation, qubits are indexed numerically. For instance,  $\texttt{qc.h(0)}$ applies a Hadamard gate to the logical qubit with index 0, denoted as $q_0$. Similarly, $\texttt{qc.cx(0, 1)}$ applies a CNOT between logical qubit $q_0$ with index 0 and $q_1$ with index 1, where $q_0$ serves as the control qubit and $q_1$ as the target qubit. A quantum program is a complete computational task that encompasses quantum circuits, while a quantum circuit is the core computational module describing qubit gate operations. A quantum circuit is a sequence of qubit gates acting on logical qubits, which are mapped to physical qubits during execution. Fig. \ref{fig-1} illustrates the quantum circuit of a decomposed Fredkin gate and its corresponding directed acyclic graph (DAG). Within the DAG,  nodes correspond to qubit gates, edges indicate causal dependencies, and the element $cx$ represents a CNOT gate. The circuit depth, defined as the length of the longest path in the DAG, reflects the number of parallelizable operation layers and is a key metric for estimating execution time and decoherence on NISQ devices \cite{arute}. 

\subsection{Superconducting quantum computer}

\begin{table*}[ht]
    \caption{Partial calibration data of $\textit{ibm-kingston}$.}
    \centering
    \begin{tabular}{cccccccc}
         \toprule
         Qubit &$T_1(\mu s)$ &$T_2(\mu s)$& \makecell{Readout assignment\\error} &ID error &RX error &Pauli-X error &CZ error \\ 
         \midrule
         0 &\num{2.796e2} &\,\,\num{3.533e2} &\num{2.881e-2} &\num{1.536e-4} &\num{1.536e-4} &\num{1.536e-4} &$0$-$1$:\num{1.111e-3} \\ 
         \midrule
         
         1 &\num{3.690e2} &\,\,\num{6.158e2} &\num{7.568e-3} &\num{2.772e-4} &\num{2.772e-4} &\num{2.772e-4} & $1$-$2$:\num{2.738e-3} \\

         & & & & & & & $0$-$1$:\num{1.111e-3} \\ 
         \midrule
         
         2 &\num{2.997e2} &\,\,\num{1.041e2} &\num{8.545e-3} &\num{1.714e-4} &\num{1.714e-4} &\num{1.714e-4} &$2$-$1$:\num{5.738e-3} \\

         & & & & & & & $2$-$3$:\num{2.516e-3} \\ 
        \midrule

         3 &\num{1.879e2} &\,\,\num{2.183e2} &\num{8.301e-3} &\num{3.920e-4} &\num{3.920e-4} &\num{3.920e-4} &$3$-$16$:\num{2.526e-3} \\

         & & & & & & & $3$-$2$:\num{2.516e-3} \\

        & & & & & & & $3$-$4$:\num{1.611e-3} \\ 
         \bottomrule
    \end{tabular}
    \label{tab-1}
\end{table*}

Superconducting quantum computers represent one of the leading platforms in quantum computing to date. Their hardware architecture is based on superconducting qubits, supported by cryogenic control electronics and microwave resonators. Superconducting qubits are primarily implemented in two-dimensional planar arrays, connected either via direct inter-qubit couplers or globally through shared bus resonators \cite{gambetta}. Quantum logic operations, including single-qubit rotations and two-qubit entangling gates, are implemented by manipulating qubit states through microwave drives \cite{kandala2}.

Superconducting quantum processors typically implement qubits as nonlinear microwave circuits whose anharmonicity is provided by Josephson junctions; on-chip capacitive/inductive couplings-often via fixed or tunable couplers-enable two-qubit gates that are activated by microwave drives or flux-parametric modulation \cite{jiaozhun}. Fig. \ref{fig-2} illustrates the architecture of $\textit{ibm-perth}$, where $Q_0$, $Q_1$, $Q_2$, $Q_3$, $Q_4$, $Q_5$, $Q_6$ are the physical qubits in the architecture.

Due to the fragility of physical qubits and their susceptibility to environmental interference, quantum circuits executed on superconducting processors are prone to several types of errors: qubit decoherence, gate errors, readout errors, and inter-qubit crosstalk \cite{sarovar,wendin}. In fact, IBM's backend calibration data shows that error rates and coherence times vary significantly across qubits, qubit couplings, and over time \cite{oliver}. Table \ref{tab-1} presents partial calibration data for the publicly accessible $\textit{ibm-kingston}$ superconducting quantum chip as of May 30, 2025. In column $\text{CZ\ error}$, the entry $0$-$1$:$\num{1.111e-3}$ denotes that the error rate of the $\text{CZ}$ gate between physical qubits $Q_0$ and $Q_1$ is $0.11\%$. Other data entries follow the same notation.

	\section{Motivation}\label{sec_XN}

\begin{figure}[ht]
    \centering
    \includegraphics[width=160pt]{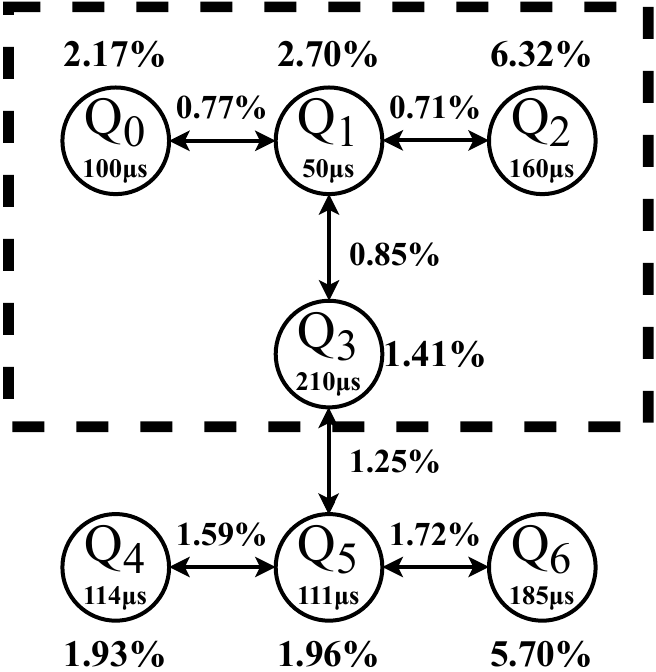}
    \caption{A case of qubit allocation. The circles represent physical qubits, with the values inside each circle indicating the qubit’s transverse relaxation time $(T_2)$. The edges between the circles denote inter-qubit coupling relationships, and each edge is annotated with the corresponding two-qubit gate error. The value placed adjacent to each physical qubit represents its readout error. Qubits within the dashed box are the subset of physical qubits designated for allocation.
    }
    \label{fig-2}
\end{figure}

In NISQ-era superconducting quantum chips, hardware limitations lead to non-uniform qubit reliabilities and heterogeneous error rates across inter-qubit couplings. The diversity of connectivity patterns and qubit counts further gives rise to distinct chip architectures \cite{kjaergaard}. To execute quantum programs on such hardware, logical qubits are typically assigned to physical ones with favorable characteristics (e.g., longer coherence times, lower gate error rates), while additional SWAP gates are inserted during execution to mediate interactions not directly supported by the native topology, thereby reconciling the program’s logical connectivity with the chip’s physical constraints. Previous noise-aware mapping techniques typically employ heuristic algorithms to identify reliable qubit mapping strategies \cite{murali2019noise,nishio2020extracting,tannu}. As illustrated in Fig. \ref{fig-2}, when executing a quantum program requiring four qubits, $Q_0$, $Q_1$, $Q_2$, and $Q_3$ are selected for their low error rates. However, these approaches only incorporate the error rates from calibration data, without accounting for the impact of the quantum program's execution time and the qubits' $T_2$ on qubit coherence.

The coherence of a qubit decays exponentially with time, while the transverse relaxation time ($T_2$) characterizes the intrinsic ability of the qubit to maintain phase coherence, and it is related to both longitudinal relaxation time ($T_1$) and pure dephasing time ($T_\phi$) \cite{nielsen}. For a single-qubit quantum state expressed in the computational basis $\{|0\rangle,|1\rangle\}$, its density matrix is $\rho=\begin{pmatrix}
\rho_{00} & \rho_{01} \\
\rho_{10} & \rho_{11}
\end{pmatrix}$,
where $\rho_{00}=|\alpha|^2$ and $\rho_{11}=|\beta|^2$ represent the probability of the qubit being in state $|0\rangle$ and $|1\rangle$, respectively. The off-diagonal elements $\rho_{01}=\alpha\beta^*$ and $\rho_{10}=\alpha^*\beta$ describe the coherence, encoding the phase relationship in the superposition state$|\psi\rangle=\alpha|0\rangle+\beta|1\rangle$. 

Pure dephasing relaxation time ($T_\phi$) characterizes phase decay independent of energy relaxation \cite{schlosshauer}. Its effect on the coherence term $\rho_{01}$ is $\rho_{01}(t)=\rho_{01}(0)e^{-t/{T_\phi}}$, where $t$ is the quantum state evolution time. Longitudinal relaxation ($T_1$) governs energy decay from $|1\rangle$ to $|0\rangle$ \cite{blum}. Its effect on the coherence term $\rho_{01}$ is $\rho_{01}(t)=\rho_{01}(0)e^{-t/{(2T_1)}}$, where $t$ is the quantum state evolution time. Combining both pure dephasing and longitudinal relaxation, the overall evolution of the coherence term $\rho_{01}$ is $\rho_{01}(t)=\rho_{01}(0)e^{-t/{T_2}}$, where $1/T_2=1/(2T_1)+1/T_\phi$. The phase coherence of a qubit thus depends on $T_2$ and evolution time $t$. Specifically, the greater the $T_2$ of a qubit and the shorter its evolution time, the stronger the coherence of the qubit.

\begin{figure}[ht]
    \centering
    \includegraphics[width=245pt]{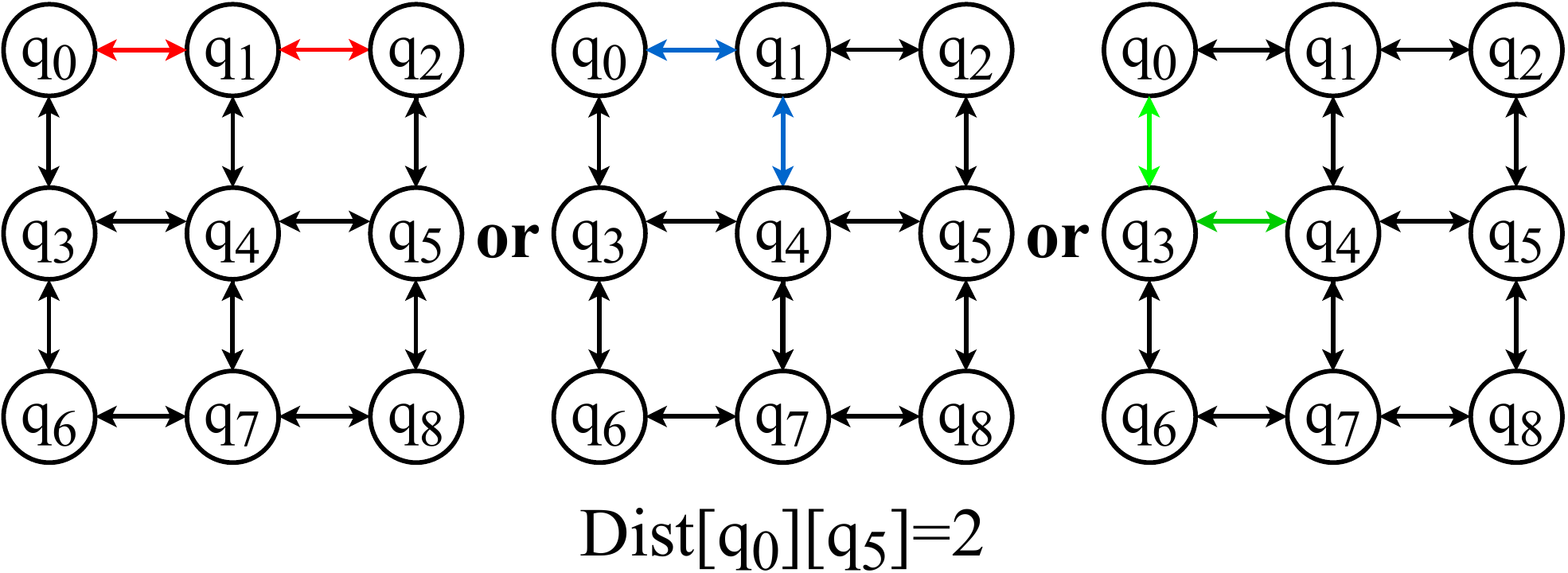}
    \caption{Different routing processes for the same quantum operations. With respect to the $\texttt{qc.cx}(0,5)$ within the same quantum program, three routing schemes based on SWAP gates are designed to mediate interactions between the logical qubits $q_0$ and $q_5$. Specifically, the red, blue, and green lines in the figure delineate the SWAP gate paths employed in each respective scheme. $\text{Dist}[q_0][q_5]=2$ indicates that the number of SWAP gates required for each scheme is $2$.}
    \label{fig-3}
\end{figure}

\begin{figure}[ht]
    \centering
    \includegraphics[width=200pt]{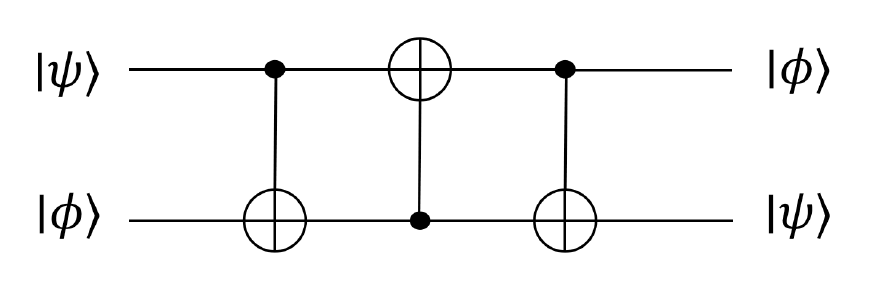}
    \caption{The quantum circuit of the decomposed SWAP gate.
}
    \label{fig-4}
\end{figure}

Similarly, in previous existing circuit compilation techniques, heuristic algorithms were commonly employed to insert SWAP gates along the shortest paths to match the hardware's physical topology. The distance between two logical qubits is defined as the minimum number of inter-qubit connections (edges in the chip's topological graph) required to link them, denoted as $\text{Dist}[q_i][q_j]$. As depicted in Fig. \ref{fig-3}, within the chip's topological graph, the minimum distance between logical qubits $q_0$ and $q_5$ is defined as $\text{Dist}[q_0][q_5]=2$, and SWAP gates are inserted along these shortest paths to adjust the positions of $q_0$ and $q_5$ to adjacent physical locations, enabling the execution of $\texttt{qc.cx}(0,5)$.

However, determining the positions of SWAP gates solely based on the minimum distance between logical qubits overlooks the practical characteristics and constraints of quantum hardware, and these factors directly impact the fidelity of circuits after SWAP gates are inserted. For instance, while the chosen path may be the shortest in distance, if it passes through physical qubits with extremely short $T_2$ and high gate error rates, it will instead lead to a decrease in overall circuit fidelity. Beyond this, SWAP gates introduce both additional noise and time overhead. As shown in Fig. \ref{fig-4}, a SWAP gate is typically decomposed into a sequence of CNOT gates, a process that accumulates errors from multiple two-qubit gates while increasing circuit depth, and thereby further affects the coherence of qubits.

\section{THE ART OF OUR DESIGN}\label{sec-TRAM}
	
\subsection{Partitioning method for superconducting quantum chips}\label{sec-CQTP}

In superconducting quantum processors of the NISQ era, the total number of calibrated physical qubits provided by the hardware typically exceeds the number of logical qubits required by a compiled quantum circuit.
This hardware redundancy enables selective mapping of logical qubits onto physically robust and well-coupled qubits, thereby improving execution fidelity, enhancing the utilization efficiency of hardware resources, and supporting strategies for parallel task execution on a single chip.

From the perspective of connectivity, mapping logical qubits onto a compact subgraph of well-connected physical qubits reduces the number of required SWAP gates. Such compact placement also mitigates routing-induced timing conflicts arising from limited parallel coupling resources, which improves gate-level parallelism~\cite{das}. Moreover, adjacent qubits are often subject to similar noise spectra; grouping them together not only facilitates consistent error mitigation strategies but also simplifies the implementation of error suppression techniques~\cite{kandala2}.

Beyond connectivity, qubit robustness is equally critical. Calibration data, such as gate and readout error rates, provide a practical measure of qubit stability. In addition, the $T_2$ captures a qubit’s resilience to decoherence. Maintaining high-fidelity entanglement relies strongly on sufficiently long and uniform $T_2$ times across the selected qubits \cite{tuixianggan,knill}. When $T_2$ values vary significantly, nonuniform coupling to the environment induces asynchronous decoherence processes \cite{wallman}, which in turn amplify crosstalk effects through qubit coupling and accelerate phase damping in entangled states, ultimately degrading the quality of entanglement \cite{arute}.

\begin{figure}[ht]
    \centering
    \includegraphics[width=245pt]{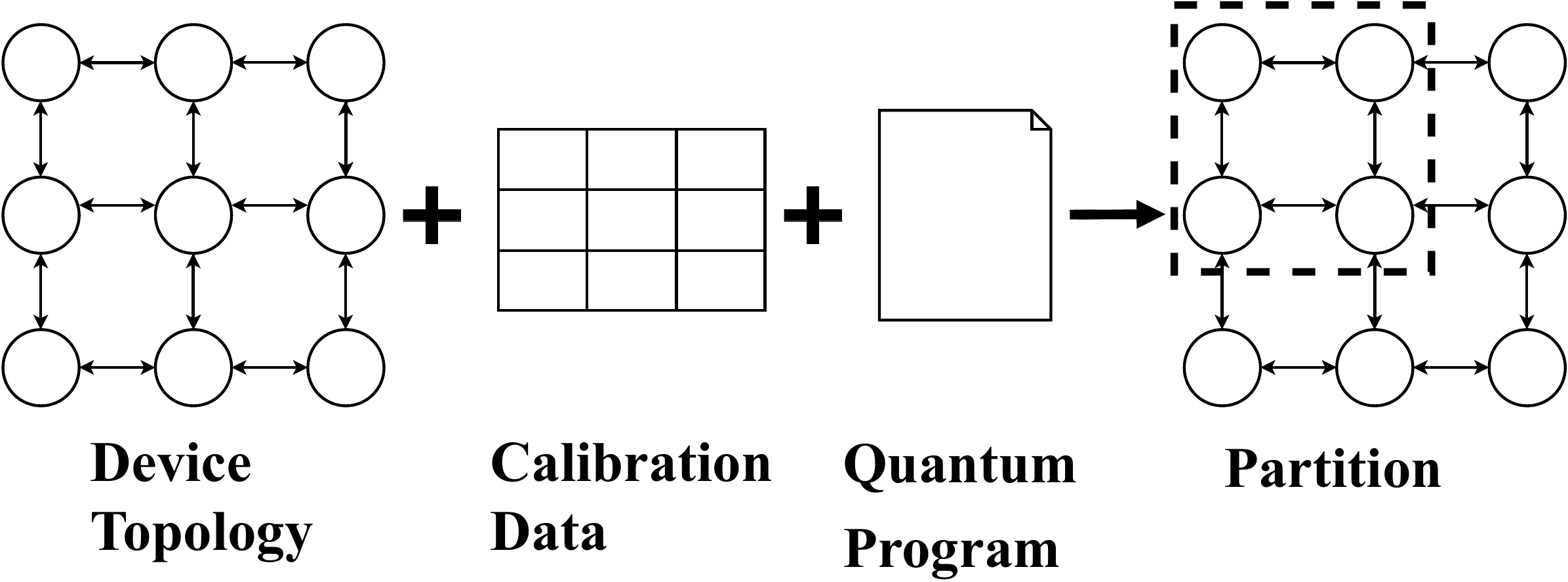}
    \caption{Qubit mapping by using CQTP in a nutshell.
    CQTP leverages hardware calibration data and the physical qubit coupling properties determined by the device topology to identify qualified schemes that comply with both the requirements of quantum programs and the physical qubit count.}
    \label{fig-5}
\end{figure}

In this paper, we propose a Community Detection-Assisted Quantum Transverse Relaxation Partitioning (CQTP) algorithm. CQTP aims to generate robust partitions of physical qubits with compact connectivity that match the logical qubit requirements of a program. 
It integrates calibration data, device topology, and transverse relaxation times of qubits to guide the selection of reliable qubit subsets that satisfy both the requirements of quantum programs and the qubit count constraint on superconducting quantum chips. As shown in Fig. \ref{fig-5}, the working principle of CQTP is demonstrated.

\begin{algorithm}[H]
\caption{FN Algorithm for Hierarchy Tree Construction}
\label{suanfa1}
\begin{algorithmic}[1]
\REQUIRE  Coupling graph $G=(V_G,E_G)$, Calibration data
\ENSURE  Hierarchy Tree $\textit{HT}$
\STATE Initialize $\textit{HT}$ as an empty tree;
\STATE Initialize each qubit as a leaf node in $\textit{HT}$;
\WHILE{there exist unmerged nodes in $\textit{HT}$}
     \STATE Initialize $F_{\max}=-\infty$, $\textit{node}_L \gets \text{None}$, and $\textit{node}_R \gets \text{None}$, where $\textit{node}_L$ and $\textit{node}_R$ denote placeholder nodes that will track the node pair with the strongest association in this iteration;
    \FORALL{pairs of unmerged nodes $(C_i, C_j)$ where $(C_i, C_j)$ are connected  in $G$, each node may be a leaf node or an internal node}
        \STATE Compute the association score $F(C_i, C_j)$;
        \IF{$F(C_i, C_j) >F_{\max}$}
            \STATE Update $F_{\max} \gets F(C_i, C_j)$, $\textit{node}_L \gets C_i$, $\textit{node}_R \gets C_j$;
        \ENDIF
    \ENDFOR
    \STATE Create a new internal node $\textit{mergedNode}$ representing the merged community of $\textit{node}_L$ and $\textit{node}_R$;
    \STATE Set $\textit{node}_L$ as the $\textbf{left subtree}$ of $\textit{mergedNode}$;
    \STATE Set $\textit{node}_R$ as the $\textbf{right subtree}$ of $\textit{mergedNode}$;
    \STATE Remove $\textit{node}_L$ and $\textit{node}_R$ from the unmerged nodes;
    \STATE Add $\textit{mergedNode}$ to $\textit{HT}$;
\ENDWHILE
\STATE \textbf{Return:} $\textit{HT}$
\end{algorithmic}
\end{algorithm}

The core principle of CQTP is to leverage community detection on the qubit coupling graph to identify clusters of physical qubits that are both strongly interconnected and exhibit comparable noise characteristics. The method constructs a hierarchical modular representation of the hardware topology using the Fast Newman (FN) community detection algorithm~\cite{newman}, which partitions the qubit network into communities with dense intra-community couplings and sparse inter-community links. This hierarchical structure provides a scalable framework for selecting subsets of qubits according to program requirements.

Given a target quantum circuit, CQTP searches this hierarchy to determine an optimal combination of communities that satisfies the logical-qubit count while maximizing overall execution fidelity. In this formulation, the choice of communities reflects a trade-off between connectivity compactness and physical robustness. Algorithm~\ref{suanfa1} details the FN-based community detection process, whereas Algorithm~\ref{suanfa2} summarizes the complete CQTP workflow from topology clustering to qubit allocation.

\begin{algorithm}[H]
\caption{CQTP Algorithm}
\label{suanfa2}
\begin{algorithmic}[1]
\REQUIRE  Calibration data $\{T_2, E, V\}$, Coupling graph $G=(V_G, E_G)$, 
Target qubit number $N_{\text{target}}$, Weight parameters $(\omega_1, \omega_2)$
\ENSURE  Valid partitions $P^*_{\text{final}}$
\STATE $\widetilde T_2 \gets {\mathrm{Normalize}({T_2, \varepsilon})}$; 
\STATE Initialize each qubit as a single community $\{Q_i\}$ and let the set of communities $\mathcal{S}$ be denoted as $\mathcal{S} \gets \{\{Q_0\}, \{Q_1\}, \ldots, \{Q_{n-1}\}\}$;

\STATE $Q_{\text{origin}} \gets {\mathrm{Modularity}(\mathcal{S}, G)}$;

\WHILE{$|\mathcal{S}| > 1$}
    \STATE Initialize $F_{\max}=-\infty$ and  $(C_i^*, C_j^*)\gets \text{None}$, where $C_i^*$, $C_j^*$ denote the connected community pair that will achieves the maximum $F$ in this iteration;

    \FORALL{$(C_i, C_j)$ \textbf{where} $(C_i, C_j)$ are connected  in $G$} 
        \STATE $Q_{\text{merged}} \gets {\mathrm{Modularity}(C_i, C_j, G)}$;
        \STATE $\Delta Q \gets {\mathrm{ModularityGain}(Q_{\text{merged}} - Q_{\text{origin}})}$;
        \STATE $T_2^{\text{SIM}} \gets {\mathrm{TemporalSimilarity}(C_i, C_j, \widetilde T_2)}$;
        \STATE $\text{EV} \gets {\mathrm{ErrorVariance}(C_i, C_j,G, E, V)}$;
        \STATE $F \gets \Delta Q + \omega_1 \times T_2^{\text{SIM}} + \omega_2 \times \text{EV}$;
        \IF{$F > F_{\max}$}
            \STATE $(F_{\max}, C_i^*, C_j^*) \gets (F, C_i, C_j)$;
        \ENDIF
    \ENDFOR
    \STATE $P_{\text{new}} \gets C_i^* \cup C_j^*$;
    \STATE $\mathcal{S} \gets (\mathcal{S} \setminus \{C_i^*, C_j^*\}) \cup \{P_{\text{new}}\}$;
    \STATE $Q_{\text{origin}} \gets {\mathrm{Modularity}(\mathcal{S}, G)}$;

    \IF{$|P_{\text{new}}| = N_{\text{target}}$}
        \STATE $P^*_{\text{final}} \gets P_{\text{new}}$;
        \STATE \textbf{break}
    \ENDIF
\ENDWHILE
\STATE \textbf{Return:} $P^*_{\text{final}}$
\end{algorithmic}
\end{algorithm}

\begin{figure}[ht]
  \centering
  \subfigure[The Process of Community Merging.]{
   \includegraphics[width=245pt]{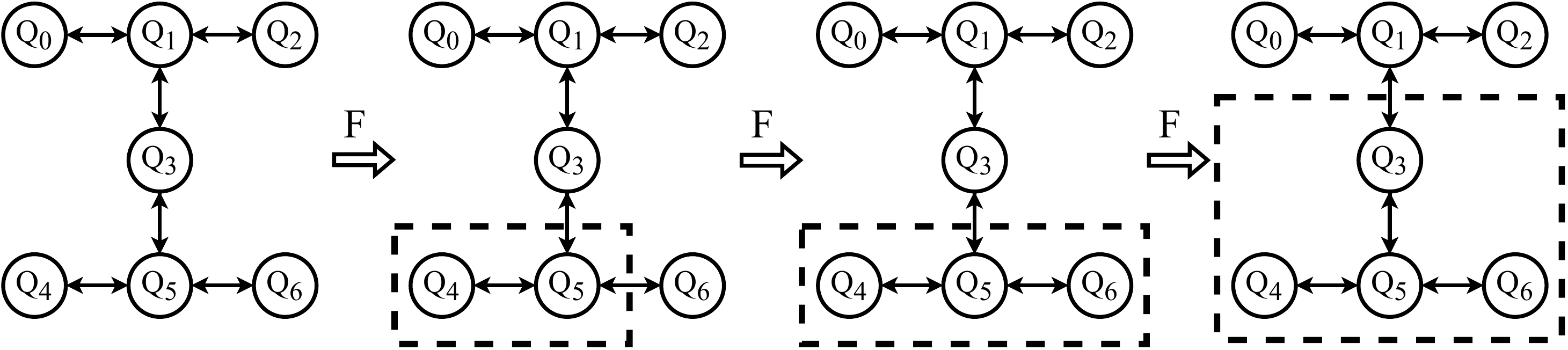}\label{fig-6(a)}
  }
  \hspace{0pt}
  \subfigure[The hierarchical tree of the community merging process.]{
   \includegraphics[width=245pt]{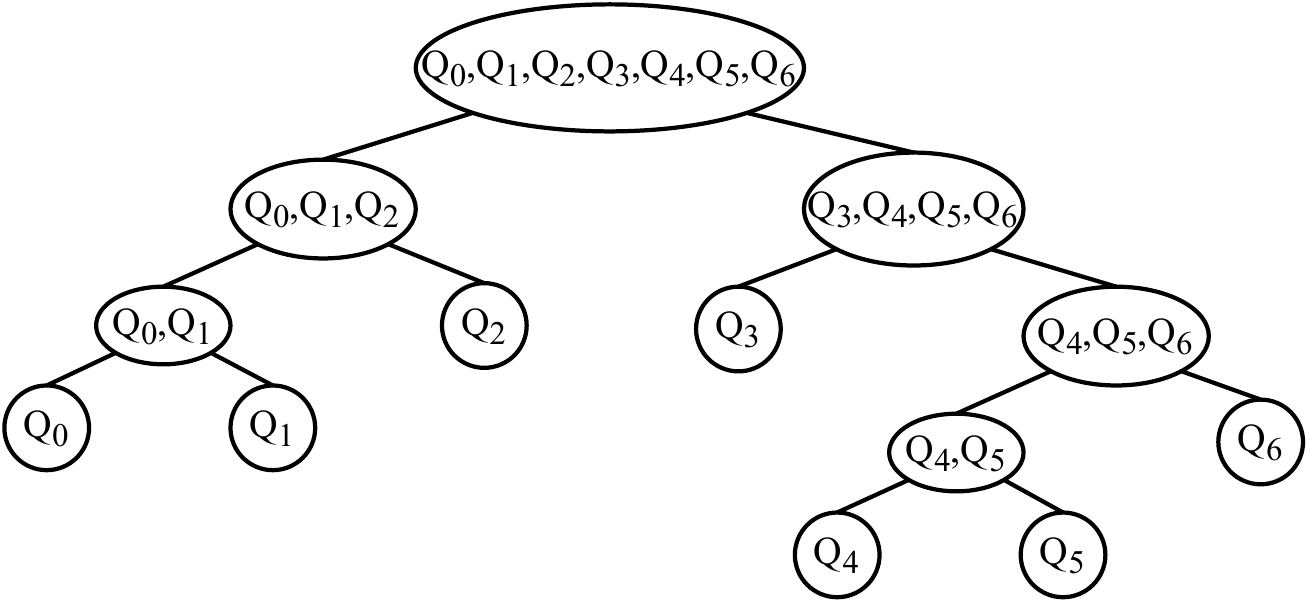}\label{fig-6(b)}
  }
  \caption{\label{fig-6} The working process of CQTP. (a) At the initial stage, each qubit is a separate community. Each time, the two communities that maximize $F$ are selected for merging. The new community formed after merging includes all qubits from the original two communities. This process continues until the new community satisfies the qubit count requirement of the quantum program. (b) The hierarchical modular representation of the hardware topology is implemented using a hierarchical tree, where each node corresponds to a modular unit of the topology.}
\end{figure}

The reward function $F$ is defined as the gain from merging two communities. In the initial stage, each physical qubit is a separate community. The CQTP algorithm calculates the reward function $F$ for all possible community merges, and each time merges the two communities that maximize the reward into a new community. This merging process continues until the number of qubits in the new community meets the qubit number requirement of the quantum program. Take a quantum program requiring four physical qubits as an example. Fig. \ref{fig-6(a)} illustrates the community merging process. In the initial stage, each physical qubit is a community and serves as a leaf node in the hierarchical tree. Each merging operation targets two existing communities rather than individual nodes, and the new community after merging includes all nodes from the original two communities. The first merge combines $Q_4$ and $Q_5$ in a new community $\{Q_4, Q_5\}$, and each subsequent merge combines the two communities that maximize the reward $F$ into a new community. The merging process continues until the count of qubits in the new community reaches four, resulting in the final community $\{Q_3, Q_4, Q_5, Q_6\}$. Fig. \ref{fig-6(b)} illustrates the the complete process of hierarchical tree of community merging.
In the initial phase, each qubit serves as an individual leaf node. Thereafter, communities that maximize the reward $F$ are merged iteratively until all qubits are consolidated into the root node of the hierarchical tree.

The form of the reward function $F$ is defined as
\begin{eqnarray}
\label{eq:F}
F = Q_{\text{merged}} - Q_{\text{origin}} + \omega_1 T_{2}^{\text{SIM}} + \omega_2 \text{EV},
\end{eqnarray}
where each term is described as follows:
\begin{itemize}
    \item \textbf{Modularity ($Q$):}  
    The modularity of a community structure is given by
    \begin{equation}
    \label{eq:Q}
        Q = \sum_i \left( e_{ii} - a_i^2 \right),
    \end{equation}
    where $e_{ii}$ denotes the fraction of edges within community $i$ relative to the total number of edges. The term $a_i = \sum_j e_{ij}$ represents the fraction of edges connected to all nodes in community $i$, with $e_{ij}$ denoting the fraction of edges between community $i$ and community $j$. A larger $Q$ value indicates denser intra-community connections and sparser inter-community connections.

    \item \textbf{Modularity gain ($\triangle Q= Q_{\text{merged}} - Q_{\text{origin}}$):}  
    Here, $Q_{\text{origin}}$ is the modularity before merging two communities, and $Q_{\text{merged}}$ is the modularity after merging them. $Q_{\text{merged}} - Q_{\text{origin}}$ quantifies the modularity gain: the larger the gain, the clearer the community structure and the stronger the topological cohesion after merging.

    \item \textbf{Temporal similarity ($T_{2}^{\text{SIM}}$):}  
    This term measures the similarity of transverse relaxation times ($T_2$) between the merged communities, capturing the extent to which the communities are physically compatible from a noise-resilience perspective.

    \item \textbf{Error variance (EV):}  
    This term reflects the variability of calibration data across the merged communities, penalizing merges that introduce higher levels of uncertainty.
\end{itemize}

In short, the reward function $F$ jointly evaluates structural modularity gain, temporal compatibility, and calibration reliability. From a physical perspective, this design ensures that the resulting partitions are not only topologically coherent but also consistent with quantum device constraints, where minimizing $T_2$ mismatches and calibration variance is critical for suppressing decoherence and improving overall execution fidelity. 

To eliminate the effect of parameter units and ensure comparability, we are required to normalize the raw $T_2$ calibration data. Meanwhile, we present the specific form of 
$T_2^{\mathrm{SIM}}$.

\begin{itemize}
    \item \textbf{Normalization.} 
    
    Let $\mathcal{S}=\{T_2( Q_i)\,|\, Q_i\, \,\text{are physical qubits}(0\leq i\leq {n-1})\}$ be the calibration set of $T_2$. We normalize $T_2$ to a unitless score $\widetilde T_2\in[0,1]$ by
    \begin{equation}
    \label{eq:t2-normalize}
    \widetilde T_2(Q_i)=\frac{T_2(Q_i)-\min \mathcal{S}}{\max \mathcal{S}-\min \mathcal{S}}.
    \end{equation}
    Since there are differences in $T_2$ across hardware qubits, ${\max \mathcal{S}-\min \mathcal{S}}\not=0$.
    For a community $C_i$, we summarize its robustness by a robust location statistic, e.g.,
    \begin{equation}
    \label{eq:t2-center}
    m_i=\operatorname{mean}\big\{\widetilde T_2(Q_i): Q_i\in C_i\big\}. 
    \end{equation}
    
    \item \textbf{Temporal similarity.} 
    
    A simple, scale-free, symmetric similarity between communities $i$ and $j$ is
    \begin{equation}
    \label{eq:t2-sim}
    T_2^{\mathrm{SIM}}
    =\frac{2\, m_i m_j}{m_i^2+m_j^2}\cdot\sqrt{\frac{m_i+m_j}{2}} \in(0,1],
    \end{equation}
    which equals $1$ iff $m_i=m_j$. 
    Specifically, $2m_im_j/(m_i^2+m_j^2)$ serves to quantify the degree of disparity between $m_i$ and $m_j$, which directly determines the baseline level of similarity. $\sqrt{(m_i+m_j)/2}$ serves to adjust the similarity value upward and further enables the selection of $m_i$ and $m_j$ with larger values and smaller disparities during the process of community merging. It is invariant to any common rescaling prior to normalization and carries no units. 
\end{itemize}

To mitigate the impact of outliers, a small constant $\varepsilon=\num{1e-8}$ is added to all $\widetilde T_2(Q_i)$. To verify that the normalization constant $\varepsilon$ does not influence the partitioning results, we conducted a robustness analysis across multiple devices ($\textit{ibm-perth}$, $\textit{ibm-guadalupe}$, $\textit{ibm-brooklyn}$) and various parameter pairs $\omega_1,\omega_2$. The partitioning outcomes remained identical for $\num{1e-10}\leq \varepsilon \leq \num{1e-6}$, which indicates that the choice of $\varepsilon=\num{1e-8}$ is numerically stable and does not affect the reward function ranking.

In addition to temporal similarity, we also incorporate hardware performance metrics to reflect the reliability of two-qubit interactions and single-qubit readouts between and within communities, respectively:  

\begin{itemize}
    \item $E$ denotes the average success rate of two-qubit gates across inter-community edges. While CNOT gates are used in our evaluation, this framework can be extended to other entangling gates such as iSWAP or CZ.
    \item $V$ denotes the average success rate of single-qubit readout operations within the two communities.
\end{itemize}

Finally, $\omega_1$ and $\omega_2$ are tunable weight parameters that balance structural modularity and physical hardware performance. Based on the empirical magnitudes of the three terms ($\Delta Q\approx\num{e-2}$ to $\num{e-1}$, $T_{2}^{\text{SIM}}\approx\num{e-1}$, and ${\text{EV}}\approx\num{e-1}$), we set $\omega_1,\omega_2\in[0,1]$ with a step size of 0.01. This range ensures that all components of the reward function remain comparable in scale, preventing any single term from dominating the optimization and allowing exploration of physically meaningful trade-offs between modularity gain and hardware reliability.

We model the objective as a weighted combination of a \textit{topology term} and two \textit{noise-aware terms} (coherence and operational errors). Let $\omega_1,\omega_2 \geq 0$ control the relative importance of coherence and error metrics, respectively. When $\omega_1=\omega_2=0$, CQTP reduces to a \textit{purely topology-driven} mapping that ignores device noise---useful as a baseline or for simulations in which noise is abstracted away. As $\omega_1$ and $\omega_2$ increase, the objective becomes \textit{noise-aware}; in the large-weight limit, topology mainly acts as a tie-breaker and the procedure approaches a \textit{noise-dominated greedy selection} guided by coherence and calibrated error data, which is appropriate for present-day NISQ hardware. 

On real devices, increasing $\omega_1$ biases CQTP toward connected subgraphs whose physical qubits exhibit \textit{long and homogeneous $T_2$ times} (e.g., large mean/minimum $T_2$ and small variance), thereby mitigating dephasing-induced information loss during execution. Increasing $\omega_2$ biases CQTP toward qubits and couplings with \textit{low calibrated operational errors}--low single-/two-qubit gate error rates and high readout fidelity--thereby reducing error accumulation and improving end-to-end circuit fidelity.

\begin{figure*}[ht] 
  \centering  
  \subfigure[Partition for multi-qubit long-entanglement workloads.]{
   \includegraphics[width=0.3\textwidth]{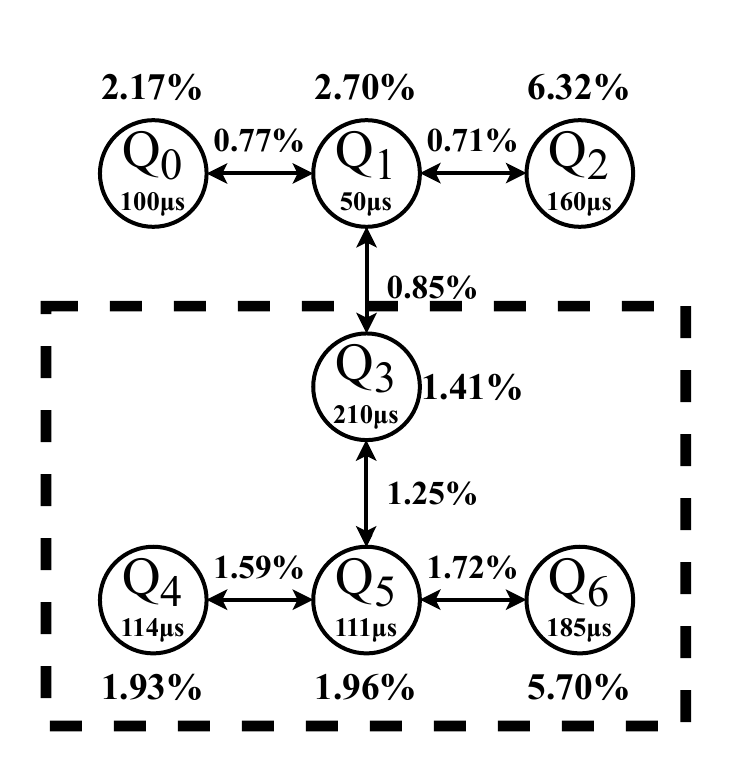}
  }
  \hfill 
  \subfigure[Partition for gate-intensive workloads.]{
   \includegraphics[width=0.3\textwidth]{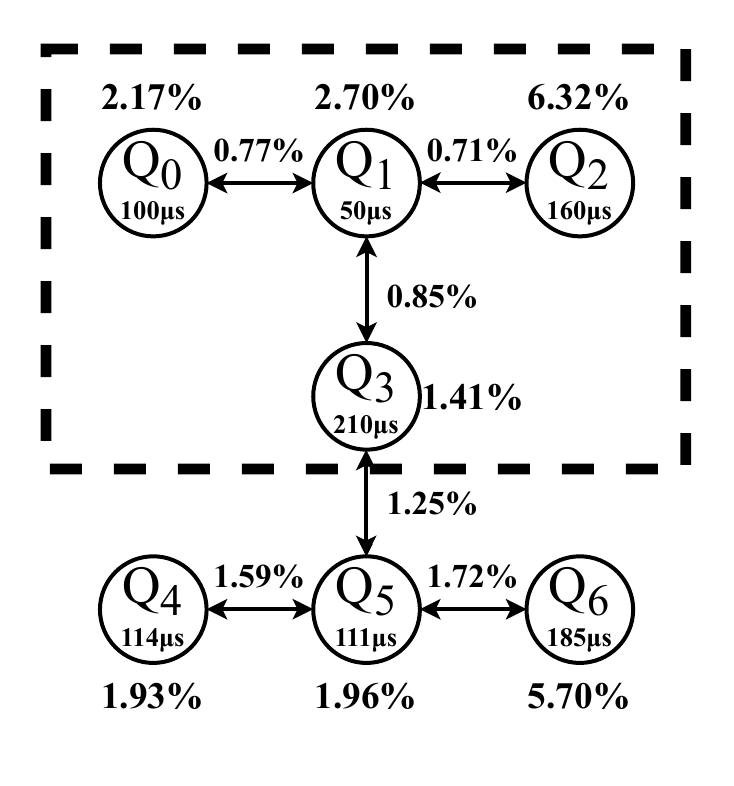}
  }
  \hfill  
  \subfigure[Partition for deep or large-scale workloads.]{
   \includegraphics[width=0.3\textwidth]{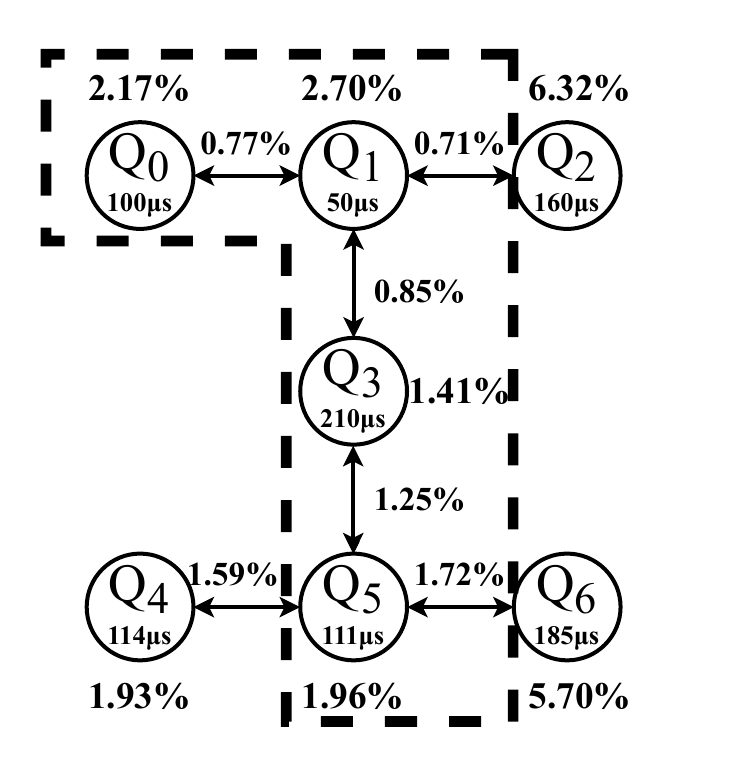}
  }
  \caption{\label{fig-7}
  Different partitioning results on the same quantum chip. Qubits within the dashed box are the subset of physical qubits designated for allocation.
  (a) In this Partition, $Q_3$, $Q_4$, $Q_5$, and $Q_6$ are selected because these qubits possess larger $T_2$ and lower dispersion, making this partition more suitable for circuits that require maintaining multi-qubit entanglement over long intervals.
  (b) In this Partition, $Q_0$, $Q_1$, $Q_2$, and $Q_3$ are selected because the two-qubit gates between adjacent qubits have lower error rates and exhibit stronger robustness, making it more suitable for gate-intensive workloads.
  (c) In this partition, $Q_0$, $Q_1$, $Q_3$, and $Q_5$ are selected because this partition achieves the simultaneous optimization of coherence performance and operational errors, and minimizes noise-induced impacts as much as possible, making it more suitable for deep or large-scale circuits.
  }
\end{figure*}

In practice, CQTP sets the coherence weight $\omega_1$ and the operational-error weight $\omega_2$ from program and calibration statistics so that topology, coherence, and error terms act on comparable scales. All objective terms need to be on the same order of magnitude so that changes in $\omega_1,\omega_2$ have interpretable and stable effects, and the weights may be scheduled across layers to mirror time-varying exposure to decoherence and error accumulation.  

The weighting parameters $\omega_1$ and $\omega_2$ are adjusted according to the operational characteristics of the target circuit. 
For circuits that must sustain multi-qubit entanglement over extended intervals---quantified by an entanglement dwell time $\tau_{\mathrm{ent}}$ \cite{PhysRevLett.93.140404} comparable to or exceeding the median $T_2$ of candidate qubits---a larger $\omega_1$ is preferred. This choice biases the optimization toward connected subgraphs with high minimum $T_2$ and low $T_2$ dispersion, mitigating dephasing-induced coherence loss. When $\tau_{\mathrm{ent}}$ is much shorter than $T_2$, the influence of coherence time diminishes and $\omega_1$ can be reduced accordingly. 

For gate-intensive workloads characterized by a high two-qubit-gate density $\rho_{2q}$, $\omega_2$ is increased to emphasize qubits and couplings with low calibrated gate and readout errors, thereby constraining stochastic error accumulation. 
When $\rho_{2q}$ is modest, maintaining a moderate $\omega_2$ keeps topological compactness as the dominant factor and minimizes additional SWAP overhead. 

In deep or large-scale circuits where both $\tau_{\mathrm{ent}}/T_2$ and $\rho_{2q}$ are elevated, $\omega_1$ and $\omega_2$ are jointly increased, accepting some reduction in geometric compactness to improve the overall success probability of circuit execution. 

As illustrated in Fig.~\ref{fig-7}, varying these parameters yields partitions aligned with different operational demands. 
Partition~(a) comprises qubits with uniformly long $T_2$, providing superior resilience against decoherence and thus supporting circuits that require stable entanglement over time. 
Partition~(b) features low two-qubit error rates across adjacent qubits, enhancing reliability under high-gate-density workloads. 
Partition~(c) achieves a balanced configuration that simultaneously optimizes coherence and gate performance, suitable for deep or noise-sensitive circuits.

\subsection{Initial mapping}\label{sec-THIM}

\begin{figure}[ht]
  \centering
  \subfigure[Initial mapping without additional routing operations.]{
   \includegraphics[width=114pt]{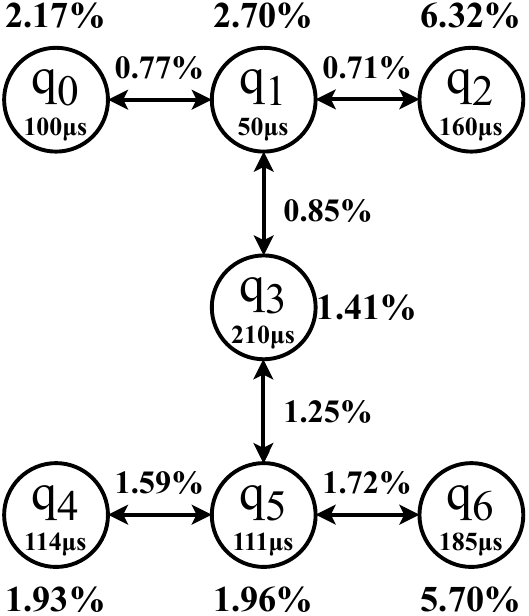}
  }
  \hspace{0pt}
  \subfigure[Initial mapping with additional routing operations.]{
   \includegraphics[width=114pt]{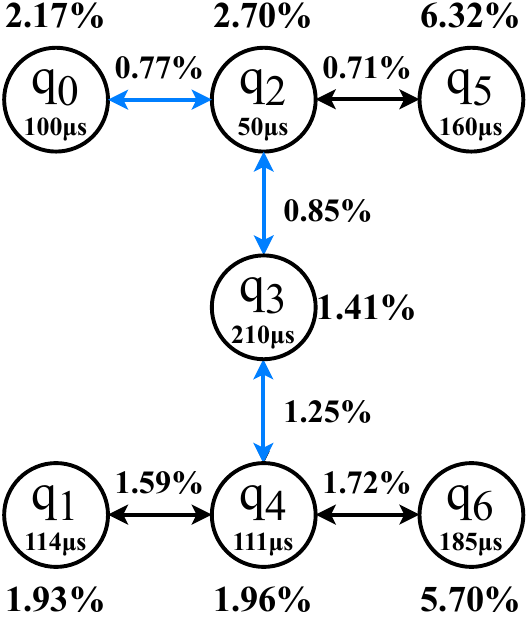}
  }
  \caption{\label{fig-8} Different routing processes of the same quantum operation $\texttt{qc.cx(0,1)}$ under different initial mapping schemes.
  (a) Under this initial mapping scheme, no additional SWAP gates are required, enabling the direct execution of quantum operation $\texttt{qc.cx(0,1)}$. (b) Under this initial mapping scheme, three additional SWAP gates are required to  realize the quantum operation $\texttt{qc.cx(0,1)}$. The blue bidirectional arrows represent SWAP gate paths. }
\end{figure}

Initial mapping refers to the starting step of assigning logical qubits to physical qubits. An optimal initial mapping can not only reduce the number of SWAP gates but also make full use of robust qubits and connections on the quantum chip \cite{kandala2}. As shown in Fig. \ref{fig-8}, for the same quantum operation $\texttt{qc.cx(0,1)}$, different initial mapping schemes have different routing processes during circuit compilation. \cite{rudinger}.

In this paper, we propose a Time-Weighted Heatmap-Based Initial Mapping (THIM) algorithm. This algorithm fully considers the qubit transverse relaxation times and the timing position of qubit gates. By adding dynamic weight allocation to two-qubit gates, it incorporates the coherence of physical qubits into the initial mapping consideration. Based on the global cost function, it achieves the optimization of the initial mapping. The THIM algorithm consists of five key stages: DAG construction and gate ordering, time-weighted heatmap generation, noise-aware routing cost calculation, greedy mapping establishment, and global cost optimization. Each stage is designed to address specific challenges in quantum compilation.

\textbf{Stage 1: DAG Construction and Topological Ordering.}
We first construct a directed acyclic graph (DAG) representation of the quantum circuit, where nodes represent qubit gates and directed edges encode gate dependencies (Fig. \ref{fig-1(b)}). To prioritize gates that are more susceptible to decoherence errors, we traverse the DAG in reverse topological order, generating an ordered set $\{TG\} = \{g_1, g_2, \ldots, g_{|TG|}\}$ of all two-qubit gates. Here, $g_1$ corresponds to the last two-qubit gate in the original circuit execution order, while $g_{|TG|}$ represents the first. This reverse ordering ensures that gates executed later in the circuit, which operate on qubits that have undergone longer coherence decay, receive higher priority in the mapping process.

\textbf{Stage 2: Time-Weighted Heatmap Construction.}
We construct an $N \times N$ symmetric heatmap matrix $\mathrm{HM}\in\mathbb{R}^{N\times N}$ (where $N$ is the number of logical qubits) that quantifies the weighted interaction frequency between logical qubit pairs. Each entry $\mathrm{HM}[q_i, q_j]$ accumulates the time-weighted contributions of all two-qubit gates acting on the logical qubit pair $[q_i,q_j]$:
\begin{equation}
\label{eq:HM}
\mathrm{HM}[q_i, q_j] = \sum_{k \in TG_{ij}} \omega_k,
\end{equation}
where $TG_{ij}$ is the index set of two-qubit gates on $[q_i,q_j]$ ordered in reverse topological time, $k=1,\dots,|TG|$. $\omega_k$ is the time weight factor for gate $g_k$. The time weight is designed to emphasize gates that execute later in the circuit:
\begin{equation}
\label{eq:wk}
\omega_k = \exp\!\left[\varphi(1-\frac{k}{|TG|})\right],
\end{equation}
where $k$ is the index of the gate $g_k$ in reverse topological order (ranging from $1$ to $|TG|$), and $\varphi \geq 0$ is an adjustable parameter that controls the degree of time-based weighting. When $\varphi = 0$, all gates receive equal weight. As $\varphi$ increases, gates later in the circuit receive exponentially higher weights. This exponential weighting scheme reflects the cumulative impact of decoherence: gates executed later operate on qubits that have experienced longer periods of relaxation and dephasing, making their placement more critical for overall circuit fidelity.

\textbf{Stage 3: Noise-Aware Routing Cost Calculation.}
Standard device calibration, e.g., randomized benchmarking (RB), reports per-gate fidelities under repeated reinitialization. 
While informative, these figures do not capture cumulative decoherence during actual execution, where qubits may remain in superposition or entangled across multiple gates~\cite{magesan1}. 
We therefore define a routing cost that combines gate-error accumulation along the coupling graph with a coherence-time dependent penalty.

\paragraph{Gate-error term.}
Let $G=(V,E)$ denote the physical coupling graph. 
For each edge $e=(Q_u,Q_v)\in E$, let $w(e)\in\mathbb{R}_{\ge 0}$ be the \emph{additive} routing cost derived from the calibrated two-qubit performance on $(Q_u,Q_v)$ (if a SWAP on $e$ is decomposed into $s$ native two-qubit gates, one may use $w_{\mathrm{SWAP}}(e)=s\,w(e)$). 
The \emph{gate-error distance} between the physical qubits $(Q_i,Q_j)$ is defined as the weighted shortest-path distance
\begin{equation}
\label{eq:Derr-shortest}
D_{\mathrm{err}}(Q_i,Q_j)\;=\;\mathrm{dist}_w(Q_i,Q_j)
\;=\;\min_{p\in\mathcal{P}(Q_i,Q_j)}\;\sum_{e\in p} w(e),
\end{equation}
where $\mathcal{P}(Q_i,Q_j)$ denotes the set of all simple paths in $G$ from $Q_i$ to $Q_j$, and $p$ is an element of $\mathcal{P}(Q_i,Q_j)$, denoting one of the simple paths from $Q_i$ to $Q_j$ in $G$. In a logarithmic-error parameterization, one can instantiate $w(e)=-\log F_{2q}(e)$ with $F_{2q}(e)\in(0,1]$ the calibrated two-qubit gate fidelity on $e$, which makes $D_{\mathrm{err}}$ strictly additive across hops and aligns with standard shortest-path solvers (e.g., Dijkstra, Floyd-Warshall).

\paragraph{Decoherence term.}
For each physical qubit $Q_i$, we model the dephasing-induced failure probability over its \emph{effective dwell time} $t_i$ as
\begin{equation}
\label{eq:PQi}
P(Q_i) \;=\; 1 - \exp\!\left[-\,\frac{t_i}{T_2(Q_i)}\right],
\end{equation}
where $T_2(Q_i)$ is obtained from calibration. 
In practice, $t_i$ can be instantiated as the qubit's scheduled wall-clock residence in superposition/entanglement during the routed interaction window; when a fine-grained schedule is unavailable, a conservative approximation $t_i\!\approx\! t$ (the circuit or layer duration relevant to the interaction) may be used.

\paragraph{Composite routing cost.}
The \emph{noise-aware routing cost} for realizing a two-qubit interaction on $(Q_i,Q_j)$ is
\begin{equation}
\label{eq:routing_cost}
D(Q_i,Q_j) \;=\; D_{\mathrm{err}}(Q_i,Q_j) \;+\; \eta\,\big(P(Q_i)+P(Q_j)\big),
\end{equation}
with $\eta\!\ge\!0$ a dimensionless weight balancing gate infidelity versus decoherence risk. 
When $\eta\!=\!0$, routing is determined solely by calibrated gate errors, favoring paths with minimal two-qubit error accumulation. 
As $\eta$ increases, the optimizer progressively prioritizes qubits with longer $T_2$ (smaller $P(\cdot)$), thereby mitigating dephasing-induced fidelity loss in time-intensive interactions.

\textbf{Stage 4: Greedy Mapping Establishment.}
We establish an initial mapping $\pi: q \rightarrow Q$ (a surjective function from logical qubits to physical qubits) using a greedy strategy based on the heatmap and routing costs:

\begin{enumerate}
    \item Sort all logical qubit pairs $[q_i, q_j]$ in descending order of their heatmap values $\mathrm{HM}[q_i, q_j]$. This prioritizes mapping frequently interacting logical pairs first.
    
    \item Sort all physical qubit pairs $(Q_i, Q_j)$ in ascending order of their routing costs $D(Q_i, Q_j)$. This creates a prioritized list of high-quality physical connections.
    
    \item Sequentially process each logical pair in sorted order. For each logical pair $[q_i, q_j]$, assign the best available physical pair $(Q_i, Q_j)$ (i.e., the physical pair with lowest routing cost among those not yet fully assigned). Update the mapping $\pi$ by setting $\pi(q_i) = Q_i$ and $\pi(q_j) = Q_j$.
\end{enumerate}

While this greedy approach efficiently handles high-frequency logical interactions, it may result in suboptimal allocation of less frequent pairs, as later assignments are constrained by earlier decisions.

\textbf{Stage 5: Global Cost Optimization.}
To overcome the limitations of greedy allocation, we employ a global cost function to evaluate and refine the initial mapping. The global cost aggregates the weighted routing costs across all logical qubit pairs:
\begin{equation}
\label{eq:global_cost}
C = \sum_{[q_i, q_j] \in S_l} \mathrm{HM}[q_i, q_j] \cdot D(\pi(q_i), \pi(q_j)),
\end{equation}
where $S_{l}$ denotes the set of all logical qubit pairs (including those with no direct interactions). This function quantifies the overall quality of mapping $\pi$ by measuring how well the physical qubit assignments align with the logical interaction patterns weighted by their frequency and timing.

Starting from the greedy solution, we apply local optimization techniques (such as iterative SWAP-based refinement or simulated annealing) to minimize $C$. In each iteration, we consider modifying the mapping by swapping physical qubit assignments and accept changes that reduce the global cost. This optimization process corrects suboptimal decisions made during greedy allocation while maintaining the computational efficiency necessary for practical compilation. 

The complete THIM algorithm is summarized in Algorithm~\ref{suanfa3}, which integrates these five stages into a unified framework for noise-aware initial qubit mapping.

\begin{algorithm}[H]
\caption{THIM Algorithm}
\label{suanfa3}
\begin{algorithmic}[1]
\REQUIRE  Logical qubits set $\{q_n\}$, Physical qubits set $\{Q_n\}$, Coupling graph $G=(V_G,E_G)$,  Quantum circuit $QC$, Calibration data $\{T_2, E\}$, Parameter $\varphi$, Decoherence weight $\eta$
\ENSURE  Initial mapping $\pi_0: \{q_n\} \rightarrow \{Q_n\}$
\STATE Construct the $DAG$ of the $QC$;
\STATE Obtain ordered two-qubit gates set in reverse $DAG$ denoted as $\{TG\}$;
\STATE Initialize $\mathrm{HM}$ as a $n\times n$ zero matrix;
\FOR{$g_k \in \{TG\}$}
    \STATE $\omega_k \gets \exp\!\left[\varphi(1-\frac{k}{|TG|})\right]$;
    \STATE Obtain the logical qubit pairs $(q_i, q_j)$ on $g_k$;
    \STATE $\mathrm{HM}[q_i,q_j] \gets \mathrm{HM}[q_i,q_j] + \omega_k$;
\ENDFOR
\FORALL{$Q_i, Q_j \in \{Q_n\}$}
    \STATE $P(Q_i) \gets\; 1 - \exp\!\left[-\,\frac{t_i}{T_2(Q_i)}\right]$;
    \STATE $D_\text{err}(Q_i,Q_j) \gets \;\min_{\pi\in\mathcal{P}(Q_i,Q_j)}\;\sum_{e\in\pi} w(e)$, where $e=(Q_i, Q_j)\in E_G$, $w(e)$ be the additive routing cost derived from the calibrated two-qubit performance on $(Q_i,Q_j)$, $\mathcal{P}(Q_i,Q_j)$ denotes the set of all simple paths in $G$ from $Q_i$ to $Q_j$;
    \STATE $D(Q_i,Q_j) \gets D_\text{err}(Q_i,Q_j) + \eta(P(Q_i)+P(Q_j))$;
\ENDFOR
\STATE Sort all logical qubit pairs $[q_i,q_j]$ in descending order of the values in $\mathrm{HM}[q_i,q_j]$;
\STATE Sort all physical qubit pairs $(Q_i,Q_j)$ in ascending order of the values in $D(Q_i,Q_j)$;
\STATE Initialize the logical-to-physical qubit mapping $\pi \gets \varnothing$;
\FOR {$[q_i,q_j]$}
    \STATE Assign the best available physical pair $(Q_i,Q_j)$;
    \STATE $\pi(q_i) \gets Q_i$, $\pi(q_j) \gets Q_j$;
\ENDFOR
\STATE $C \gets \sum_{[q_i,q_j]} HM[q_i,q_j] \cdot D(\pi(q_i),\pi(q_j))$;
\STATE Optimize the initial mapping by minimizing $C$, $\pi_0\gets\pi$;
\STATE \textbf{Return:} $\pi_0$
\end{algorithmic}
\end{algorithm}

\subsection{Time Adaptive Dynamic SWAP Algorithm}\label{sec-TSWAP}

The SWAP gate is the standard mechanism for reconciling logical two-qubit gates with the limited connectivity of quantum hardware, but each SWAP decomposes into multiple CNOT gates and thus incurs additional depth, latency, and error accumulation \cite{harrigan}. If physically fragile qubits (e.g., with short $T_2$ or high error rates) are repeatedly chosen as intermediaries, the resulting depolarizing and decoherence errors compound and can dominate the fidelity loss of the compiled circuit. We refer to this phenomenon as SWAP hotspotting, where a small set of qubits are disproportionately reused as routing paths, creating localized error accumulation \cite{ayanzadeh2023frozenqubits,rached2024spatio}.  

The SABRE algorithm \cite{sabre} addresses this routing challenge using a heuristic based on the Nearest Neighbor Cost (NNC), defined as the sum of shortest-path distances between logical qubits of ready two-qubit gates under the current mapping. Candidate SWAPs are evaluated by their immediate effect on reducing this cost, and the SWAPs that most improve NNC is chosen for execution.  

Building on this idea, we propose a Time-Adaptive Dynamic SWAP (T-SWAP) algorithm, which augments the NNC heuristic with noise-awareness. In addition to minimizing routing distance, T-SWAP incorporates qubit-dependent attenuation factors that dynamically track cumulative decoherence exposure, thereby penalizing hotspot qubits and promoting more balanced routing. This discourages SWAP hotspotting, favors qubits with stronger noise resilience, and ultimately reduces both cumulative error rates and depth overhead. A formal definition of the cost function and the full T-SWAP algorithm are given in the following.

The heuristic cost function of T-SWAP is defined as:
\begin{equation}
\begin{aligned}
H =& \max\!\big( \mathrm{decay}(q_i),\, \mathrm{decay}(q_j) \big) \cdot\\& 
\Bigg[
    \frac{1}{|F_{\text{2q}}|} \sum_{p_{g} \in F_{\text{2q}}} D(Q_i,Q_j) 
    +  
    \frac{\mu}{|E_{\text{2q}}|} \sum_{p_{g} \in E_{\text{2q}}} D(Q_i,Q_j)
\Bigg],
\label{eq:tSWAP}
\end{aligned}
\end{equation}
where \(D(Q_i,Q_j)\) is the routing distance between physical qubits \(Q_i\) and \(Q_j\) on the coupling graph (Eq.~\ref{eq:routing_cost}), $p_g$ denotes the two-qubit gate in $p$ (Eq.~\ref{eq:Derr-shortest}). \(\mathrm{decay}(q)\) denotes the attenuation factor associated with logical qubit \(q\) and models the gradual loss of coherence availability for heavily used qubits, penalizing those that have recently participated in SWAP operations and thereby mitigating hotspotting. After each SWAP involving qubit \(q\), the decay factor is updated as 
\(\mathrm{decay}(q) \leftarrow \mathrm{decay}(q) + \delta\), where \(\delta > 0\) controls the rate of penalty accumulation. To achieve a dynamic balance between suppressing hotspots and controlling depth, the value selection of $\delta$ should be deeply coupled with hardware noise characteristics, circuit structure, and algorithm phases. In this paper, we select $\delta=\num{1e-3}$. The sets \(F_{\text{2q}}\) and \(E_{\text{2q}}\) correspond to the two-qubit gates in the front layer and the look-ahead layer of the circuit DAG, respectively, with \(F_{\text{2q}}\) having execution priority. The coefficient \(\mu \in [0,1]\) controls the influence of look-ahead gates in the cost function (Eq.~\ref{eq:tSWAP}), ensuring that the current layer is prioritized while still accounting for upcoming constraints: smaller values of \(\mu\) emphasize immediate execution efficiency, while larger values improve long-term routing smoothness by accounting for future dependencies.

This formulation enables T-SWAP to select SWAP operations that simultaneously reduce near-term routing cost, avoid repeated qubit reuse, and maintain consistency with upcoming connectivity constraints. Intuitively, the cost function $H$ favors SWAP gates that (i) minimize routing distance, and (ii) avoid repeatedly involving the same logical qubits. This design encourages greater parallelism and effectively reduces the overall circuit depth.

\begin{figure}[H]
  \centering
  \subfigure[The representation of $F_{2q}$ and $E_{2q}$ in quantum circuits.]{
   \includegraphics[width=245pt]{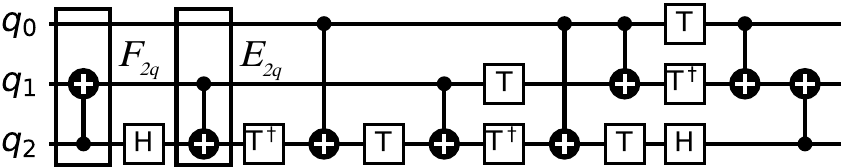}
  }
  \hspace{10pt}
  \subfigure[The representation of $F_{2q}$ and $E_{2q}$ in DAG.]{
   \includegraphics[width=245pt]{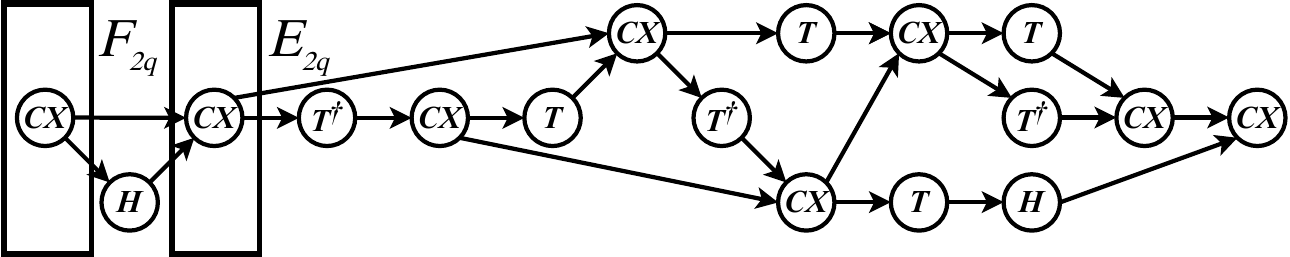}
  }
  \caption{\label{fig-9}The representation of $F_{2q}$ and $E_{2q}$ in a quantum circuit and its corresponding DAG. (a) The quantum circuit of the decomposed Fredkin gate, where $F_{2q}$ represents the current-layer two-qubit gate set and $E_{2q}$ represents the look-ahead two-qubit gate set. (b) The case in the DAG corresponding to the above quantum circuit.}
\end{figure}

We first define the relevant gate sets used in T-SWAP. The current layer \(F\) consists of all gates in the circuit DAG that have no unexecuted predecessors, and thus can be executed immediately. Gates within the same layer can be executed in parallel. Since single-qubit gates can always be executed without routing, we restrict attention to the two-qubit subset
\begin{equation*}
  \begin{aligned}
   F_{2q} = \{\, g \in F \mid g \text{ is a two-qubit gate}\,\}.
  \end{aligned}
\end{equation*}
To anticipate future dependencies, we also define the look-ahead set \(E\), which contains the immediate successors of gates in \(F\). Its two-qubit subset is denoted as
\begin{equation*}
  \begin{aligned}
   E_{2q} = \{\, g \in E \mid g \text{ is a two-qubit gate}\,\}.
  \end{aligned}
\end{equation*}
As illustrated in Fig.~\ref{fig-9}, \(F_{2q}\) corresponds to the current-layer two-qubit gate set and $E_{2q}$ represents the look-ahead two-qubit gate set.

The T-SWAP procedure can then be summarized as follows:
\begin{enumerate}
    \item \textbf{Initialization:} Form the initial front layer \(F_{2q}\) by selecting two-qubit gates with no unexecuted predecessors.
    \item \textbf{Execution check:} For each gate in \(F_{2q}\), test whether it can be executed directly under the current logical-to-physical mapping. Executable gates are moved into the set \(M\) of ready gates, executed, and removed from \(F_{2q}\). Their successors are then promoted into \(F_{2q}\) once all their dependencies are resolved.
    \item \textbf{Routing by SWAP:} If no gate in \(F_{2q}\) is executable (i.e. $M = \varnothing$), candidate SWAPs are generated by considering neighbors of the mapped physical qubits in the coupling graph. Each candidate is scored using the T-SWAP cost function \(H\) (Eq.~\ref{eq:tSWAP}). The SWAP with the minimum cost is selected, the mapping is updated, and the process continues.
    \item \textbf{Iteration:} Steps (2)–(3) are repeated until \(F_{2q}\) becomes empty, meaning all two-qubit gates have been scheduled. The resulting mapping is denoted \(\pi_f\).
\end{enumerate}

In essence, \(F_{2q}\) represents the gates that can be executed now, while \(E_{2q}\) provides a limited look-ahead. The T-SWAP routine alternates between executing available two-qubit gates and applying carefully chosen SWAPs to make further progress. This design balances immediate execution with future readiness, reducing overall circuit depth. Algorithm \ref{suanfa4} demonstrates the process of T-SWAP.

\begin{algorithm}[H]
\caption{T-SWAP Algorithm}
\label{suanfa4}
\begin{algorithmic}[1]
\REQUIRE DAG, Initial mapping $\pi_0$, Routing cost $D[Q_i,Q_j]$, Decay increment $\delta$, Look-ahead weight $\mu$
\ENSURE  Final mapping $\pi_f$

\STATE Let $F_{2q}$ be denoted as the two-qubit gates with no unexecuted predecessors in DAG;
\STATE Let $E_{2q}$ be denoted as the immediate successors of gates in $F_{2q}$;
\STATE $\pi \gets \pi_0$;
\STATE $\text{decay}[q] \gets 1$ for all logical qubits $q$;
\WHILE{$F_{2q} \neq \varnothing$}
    \STATE $M \gets \varnothing$, $M$ is ready set of executable two-qubit gates;
    
    \FOR{$g \in F_{2q}$}
        \IF{$g$ is executable in $\pi$}
            \STATE $M \gets M \cup \{g\}$;
        \ENDIF
    \ENDFOR
    
    \IF{$M \neq \varnothing$}
        \FOR{$g \in M$}
            \STATE Execute $g$;
            \STATE $F_{2q} \gets F_{2q} \setminus \{g\}$;
            \STATE $F_{2q} \gets F_{2q} \cup \{g'\}$, $\{g'\}$ is the successors ready gate of $g$ in $E_{2q}$;
        \ENDFOR
    \ELSE
        \STATE Generate the set of candidate SWAP gates: $\textit{SWAPCandidates}$ based on mapping $\pi$;
        \FOR{$\textit{SWAP}\in \textit{SWAPCandidates}$}
           \STATE Based on $\textit{SWAP},F_{2q},E_{2q},D,\text{decay}$ and $\mu$ to compute $H(\textit{SWAP})$;
        \ENDFOR
        \STATE Obtain the $\textit{BestSWAP}\in\textit{SWAPCandidates}$ that minimizes $H$;
        \STATE Update mapping $\pi$ based on $\textit{BestSWAP}$;
        \STATE $\text{decay}[\textit{BestSWAP}.q_i] \gets \text{decay}[\textit{BestSWAP}.q_i] + \delta$;
        \STATE $\text{decay}[\textit{BestSWAP}.q_j] \gets \text{decay}[\textit{BestSWAP}.q_j] + \delta$;
    \ENDIF
\ENDWHILE
\STATE $\pi_f \gets \pi$;
\STATE \textbf{Return:} $\pi_f$
\end{algorithmic}
\end{algorithm}
	\section{Evaluation}\label{sec-EVAL}
     \subsection{Evaluation metrics}

\begin{table*}
    \caption{Simulators and Benchmarks.}
\centering
\begin{tabular}{|c|c|}
\hline
Simulator & Benchmark \\
\hline
\textit{ibm-perth} & \makecell{$\textit{dnn\_n2}$, $\textit{deutsch\_n2}$, $\textit{quantumwalks\_n2}$,\\  $\textit{basis\_change\_n3}$, $\textit{fredkin\_n3}$, $\textit{linearsolver\_n3}$} \\
\hline
\textit{ibm-guadalupe} & \makecell{$\textit{basis\_trotter\_n4}$, $\textit{bell\_n4}$, $\textit{variational\_n4}$, $\textit{vqe\_n4}$, $\textit{hs4\_n4}$,\\
$\textit{error\_correction3\_n5}$, $\textit{qaoa\_n6}$, $\textit{dnn\_n8}$, $\textit{hhl\_n7}$, $\textit{vqe\_ucces\_n4}$,\\
$\textit{vqe\_ucces\_n6}$, $\textit{vqe\_ucces\_n8}$} \\
\hline
\textit{ibm-brooklyn}& \makecell{$\textit{square\_root\_n18}$, $\textit{qft\_n26}$, $\textit{qv\_n32}$, $\textit{multiplier\_n45}$, $\textit{dnn\_n16}$,\\
$\textit{add\_n28}$, $\textit{wstate\_n27}$,
$\textit{is\_n34}$} \\
\hline
\end{tabular}
    \label{tab-2}
\end{table*}

(1) Fidelity of quantum states: In quantum simulators, we utilize quantum state tomography based on MLE to reconstruct the density matrix and calculate the fidelity of quantum states. To accurately reconstruct the density matrix, we compile and run each quantum benchmark on the quantum simulator for 10000 trials.

(2) The number of compiled qubit gates: We evaluate the compiler's capability to reduce compilation overhead during quantum program compilation by the number of compiled qubit gates.

(3) Compiled circuit depth: The circuit depth of the compiled quantum program is used to assess the compiler's ability to reduce coherent errors.

     \subsection{Quantum simulator}
\begin{figure}[ht]
  \centering
  \subfigure[Architecture of $\textit{ibm-perth}$.]{
    \includegraphics[width=110pt]{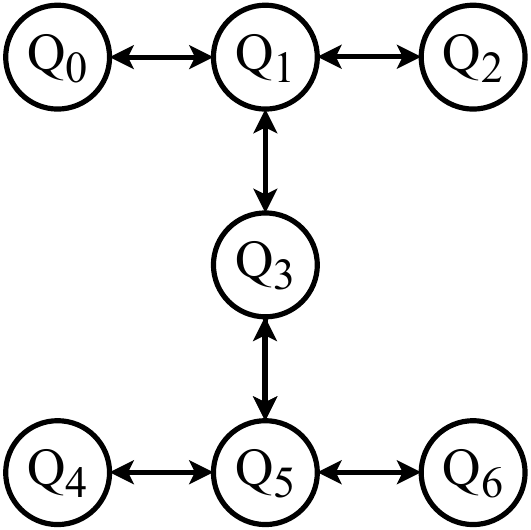}\label{fig-10(a)}
  }
  \hspace{0pt}
  \subfigure[Architecture of $\textit{ibm-guadalupe}$.]{
   \includegraphics[width=119pt]{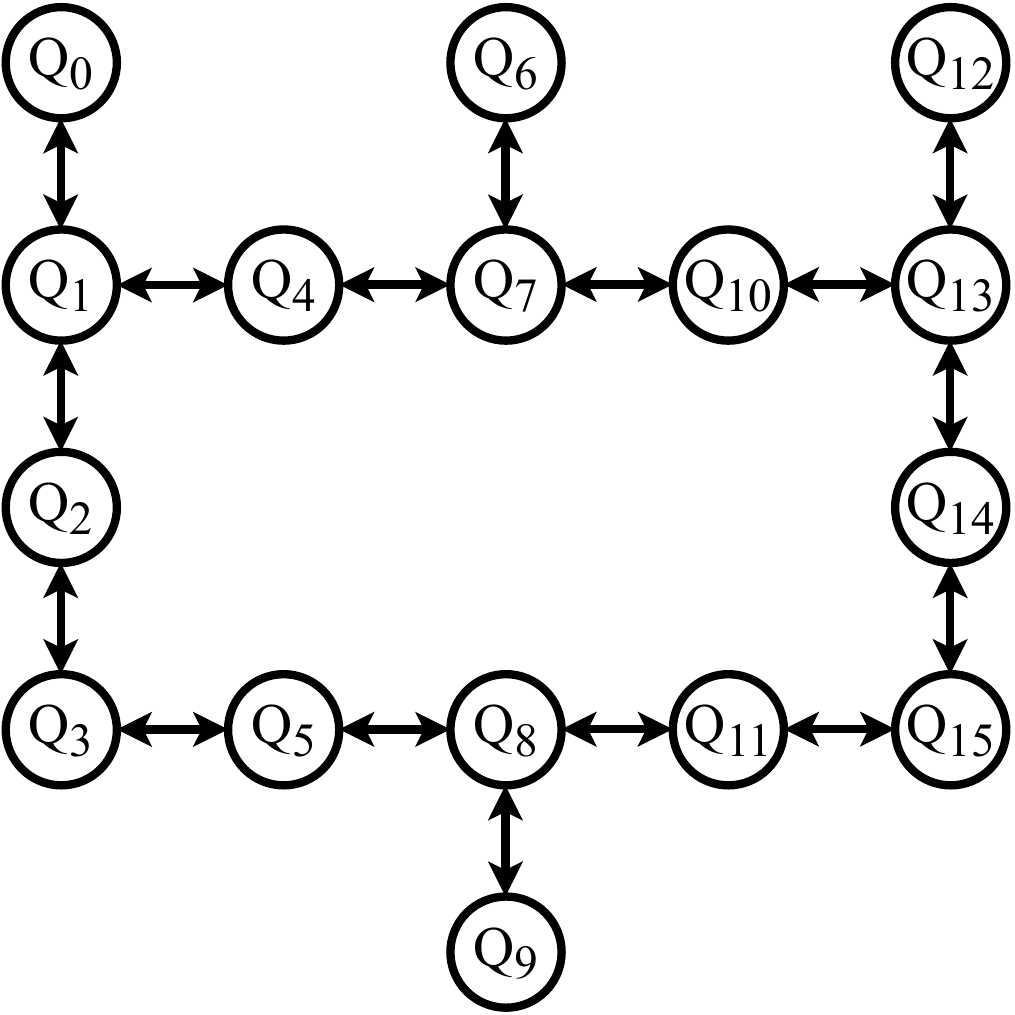}
  }
\subfigure[Architecture of $\textit{ibm-brooklyn}$.]{
   \includegraphics[width=240pt]{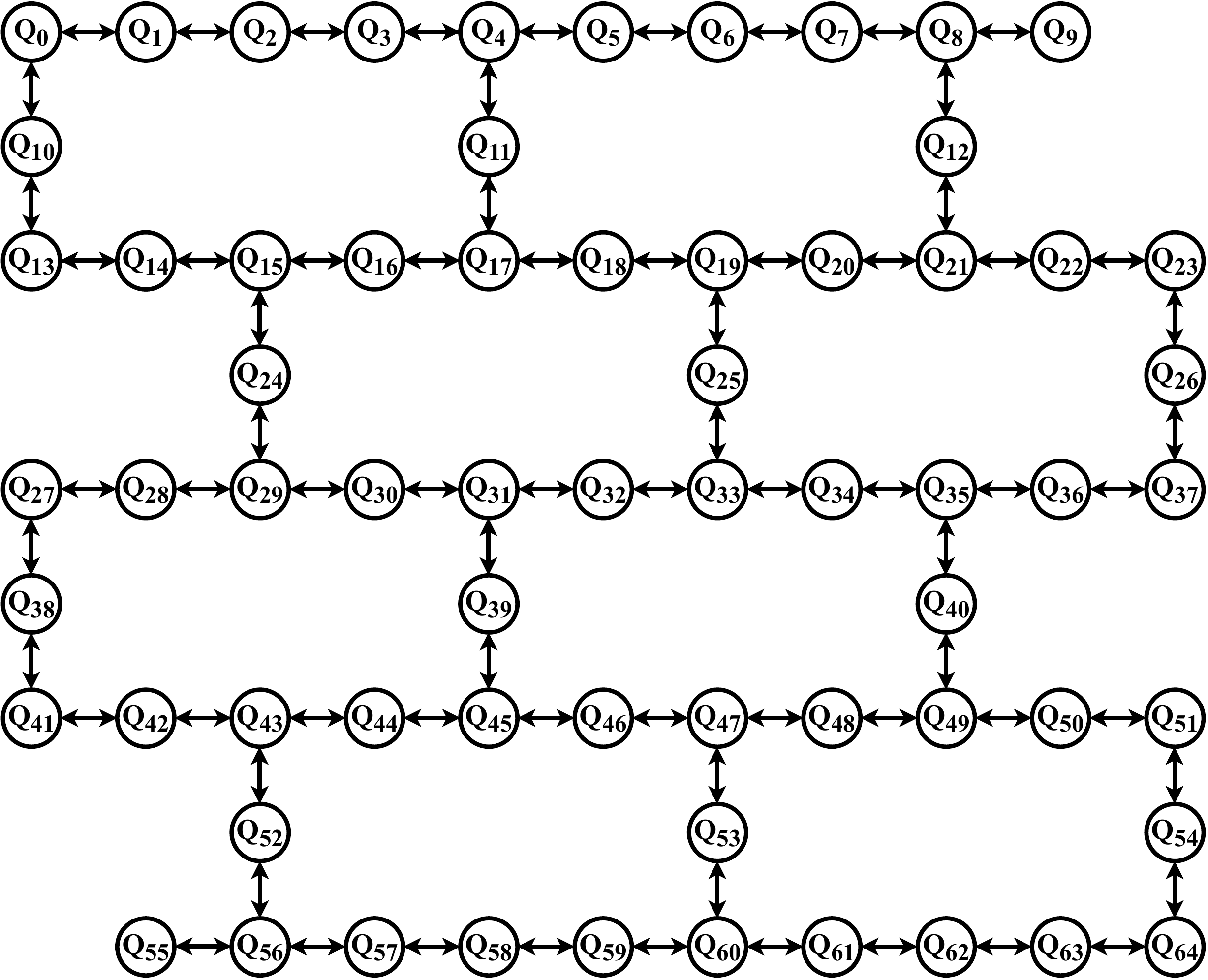}
  }
  
  \caption{\label{fig-10} Topologies of IBM QPUs. (a) The 7-qubit $\textit{ibm-perth}$. (b) The 16-qubit $\textit{ibm-guadalupe}$. (c) The 65-qubit $\textit{ibm-brooklyn}$.}
\end{figure}

We evaluate our work on quantum simulators. We employ the Qiskit framework to develop quantum simulators that reproduce the physical characteristics of real quantum hardware by incorporating depolarizing, dephasing, and amplitude damping noise models. The noise models are implemented in operator form, and the simulator’s topological connectivity and calibration parameters are derived from real IBM quantum hardware data. The quantum backends employed in this study include $\textit{ibm-perth}$ \cite{pelofske}, $\textit{ibm-guadalupe}$ \cite{Ramadhani}, and $\textit{ibm-brooklyn}$ \cite{pelofske}, whose architectures are illustrated in Fig.~\ref{fig-10}.

 \subsection{Benchmarking}

The benchmarks we used are derived from the examples in QASMBench as shown in the Tab. \ref{tab-2}. The specific quantum tasks and their corresponding simulators are shown in the table. To demonstrate the effectiveness of TRAM, the optimization level used in circuit compilation is set to 0. Only the necessary routing processes are performed to ensure that the circuit is executable on the target quantum device, with no optimizations conducted at either the logical or physical level.

 \subsection{Evaluation results}

 \begin{table*}[ht]
 \caption{This table compares the fidelity between SABRE and TRAM. The improvement quantifies the reduction in qubit gate count and circuit depth, as well as the increase in fidelity, achieved by TRAM relative to SABRE.}
\centering
\resizebox{\textwidth}{!}{
\begin{tabular}{|c|c|ccc|ccc|ccc|}
\hline
\multirow{2}{*}{Benchmark} & \multirow{2}{*}{Simulator} & \multicolumn{3}{c|}{SABRE} & \multicolumn{3}{c|}{TRAM} & \multicolumn{3}{c|}{Improvement} \\ \cline{3-11} 
 &  & Gates & Depth & Fidelity & Gates & Depth & Fidelity & Gates & Depth & Fidelity\\ \hline
\textit{dnn\_n2} & \textit{ibm-perth} & 810 & 498 & 96.48\% & 810 & 498 & 98.16\% & 0 & 0 & 1.74\% \\ \hline
\textit{deutsch\_n2} & \textit{ibm-perth} & 11 & 8 & 95.32\% & 11 & 8 & 98.13\% & 0 & 0 & 2.81\% \\ \hline
\textit{quantumwalks\_n2} & \textit{ibm-perth} & 43 & 23 & 94.09\% & 43 & 23 & 97.29\% & 0 & 0 & 3.20\% \\ \hline
\textit{basis\_change\_n3} & \textit{ibm-perth} & 185 & 113 & 88.81\% & 188 & 119 & 91.17\% & -3 & -6 & 2.66\% \\ \hline
\textit{fredkin\_n3} & \textit{ibm-perth} & 32 & 27 & 89.14\% & 32 & 27 & 94.68\% & 0 & 0 & 6.21\% \\ \hline
\textit{linearsolver\_n3} & \textit{ibm-perth} & 39 & 23 & 93.71\% & 39 & 23 & 97.31\% & 0 & 0 & 3.84\% \\ \hline
\textit{basis\_trotter\_n4} & \textit{ibm-guadalupe} & 138 & 84 & 85.68\% & 147 & 109 & 90.83\% & -9 & -25 & 6.01\% \\ \hline
\textit{variational\_n4} & \textit{ibm-guadalupe} & 70 & 45 & 90.66\% & 76 & 62 & 90.92\% & -6 & -17 & 0.29\% \\ \hline
\textit{vqe\_n4} & \textit{ibm-guadalupe} & 89 & 27 & 95.53\% & 89 & 27 & 97.27\% & 0 & 0 & 1.82\% \\ \hline
\textit{bell\_n4} & \textit{ibm-guadalupe} & 123 & 43 & 93.73\% & 123 & 43 & 96.96\% & 0 & 0 & 3.45\% \\ \hline
\textit{hs4\_n4} & \textit{ibm-guadalupe} & 68 & 23 & 92.05\% & 105 & 36 & 97.18\% & -37 & -13 & 5.58\% \\ \hline
\textit{error\_correction3\_n5} & \textit{ibm-guadalupe} & 249 & 143 & 86.42\% & 258 & 148 & 91.19\% & -9 & -5 & 5.52\% \\ \hline
\end{tabular}
}
\label{tab-3}
\end{table*}

{
\setlength{\tabcolsep}{10pt}
\begin{table*}[ht]
\caption{This table compares the compilation overhead (qubit gate count and circuit depth) between SABRE and TRAM. The improvement quantifies the percentage reduction in qubit gate count and circuit depth of TRAM relative to SABRE.}
\centering
\resizebox{\textwidth}{!}{
\begin{tabular}{|c|c|cc|cc|cc|}
\hline
\multirow{2}{*}{Benchmark} & \multirow{2}{*}{Simulator} & \multicolumn{2}{c|}{SABRE} & \multicolumn{2}{c|}{TRAM} & \multicolumn{2}{c|}{Improvement} \\ \cline{3-8} 
 &  & Gates & Depth & Gates & Depth & Gates & Depth\\ \hline

\textit{vqe\_ucces\_n4} & \textit{ibm-guadalupe} & 388 & 213 & 346 & 194 & 10.83\% & 8.92\% \\ \hline
\textit{qaoa\_n6} & \textit{ibm-guadalupe} & 936 & 439 & 792 & 391 & 15.38\% & 10.93\% \\ \hline
\textit{vqe\_ucces\_n6} & \textit{ibm-guadalupe} & 3986 & 2058 & 3689 & 1874 & 7.48\% & 8.94\% \\ \hline
\textit{hhl\_n7} & \textit{ibm-guadalupe} & 1476 & 1300 & 1146 & 1014 & 22.63\% & 22.00\% \\ \hline
\textit{vqe\_ucces\_n8} & \textit{ibm-guadalupe} & 18896 & 10193 & 18416 & 9677 & 2.54\% & 5.06\% \\ \hline
\textit{dnn\_n8} & \textit{ibm-guadalupe} & 3683 & 1100 & 3115 & 902 & 15.42\% & 18.00\% \\ \hline
\textit{dnn\_n16} & \textit{ibm-brooklyn} & 7336 & 931 & 6258 & 837 & 14.69\% & 10.09\% \\ \hline
\textit{square\_root\_n18} & \textit{ibm-brooklyn} & 8491 & 8572 & 7446 & 4893 & 12.31\% & 7.19\% \\ \hline
\textit{qft\_n26} & \textit{ibm-brooklyn} & 8255 & 1844 & 7282 & 1654 & 11.79\% & 10.30\% \\ \hline
\textit{wstate\_n27} & \textit{ibm-brooklyn} & 538 & 276 & 479 & 252 & 10.97\% & 8.70\% \\ \hline
\textit{add\_n28} & \textit{ibm-brooklyn} & 1432 & 753 & 1246 & 681 & 12.99\% & 9.56\% \\ \hline
\textit{qv\_n32} & \textit{ibm-brooklyn} & 26201 & 4019 & 22933 & 3646 & 12.47\% & 9.28\% \\ \hline
\textit{is\_n34} & \textit{ibm-brooklyn} & 3126 & 922 & 3078 & 965 & 1.54\% & 24.60\% \\ \hline
\textit{multiplier\_n45} & \textit{ibm-brooklyn} & 23672 & 10956 & 21341 & 8941 & 9.80\% & 18.39\% \\ \hline
\end{tabular}
}
\label{tab-4}
\end{table*}
}

\medskip
\noindent\textbf{(a) Fidelity evaluation and analysis.}  

We evaluate circuit fidelity using the benchmark tasks and corresponding simulators listed in Tab.~\ref{tab-2}. 
All simulators share identical calibration data for fair comparison. The results in Tab.~\ref{tab-3} demonstrate a consistent improvement in fidelity achieved by TRAM over SABRE across both backends. 

On $\textit{ibm-perth}$, TRAM attains fidelities of $98.16\%$, $98.13\%$, $97.29\%$, $91.17\%$, $94.68\%$, and $97.31\%$ for the six tested tasks, exceeding SABRE by $1.74\%$, $2.81\%$, $3.20\%$, $2.66\%$, $6.21\%$, and $3.84\%$, respectively. 
Similarly, on $\textit{ibm-guadalupe}$, TRAM achieves fidelities of $90.83\%$, $90.92\%$, $97.27\%$, $96.96\%$, $97.18\%$, and $91.19\%$, surpassing SABRE by $6.01\%$, $0.29\%$, $1.82\%$, $3.45\%$, $5.58\%$, and $5.52\%$. Averaged over all benchmarks, TRAM yields a fidelity gain of $3.59\%$, with the maximum improvement reaching $6.21\%$.

Beyond the numerical gains, TRAM exhibits a stable and noise-adaptive partitioning behavior that translates directly into higher circuit fidelity. 
In $\textit{dnn\_n2}$, $\textit{deutsch\_n2}$, $\textit{quantumwalks\_n2}$, $\textit{fredkin\_n3}$, $\textit{linearsolver\_n3}$, $\textit{vqe\_n4}$, and $\textit{bell\_n4}$, TRAM preserves both the gate count and circuit depth of SABRE, yet CQTP identifies qubit partitions with lower intrinsic error rates, thereby enhancing overall fidelity. Notably, in $\textit{dnn\_n2}$, $\textit{deutsch\_n2}$, and $\textit{quantumwalks\_n2}$, where no additional SWAP gates are required to mediate interactions between qubits, the $1.74\%$, $2.81\%$, and $3.20\%$ improvements arise from selecting a subset of more robust qubits with longer $T_2$ coherence times and implementing an optimized initial mapping.

For more entanglement-intensive circuits such as $\textit{basis\_change\_n3}$, $\textit{basis\_trotter\_n4}$, $\textit{variational\_n4}$, $\textit{hs4\_n4}$, and $\textit{error\_correction3\_n5}$, 
CQTP reduces depolarization by identifying low-error partitions, while THIM and T-SWAP jointly mitigate decoherence. TRAM integrates both mechanisms, achieving a balanced trade-off between connectivity and coherence that leads to superior fidelity across diverse workloads.

Overall, these results highlight TRAM’s ability to exploit calibration-aware qubit selection and noise-adaptive routing 
to deliver systematically higher execution fidelity without additional compilation overhead.

\medskip
\noindent\textbf{(b) Evaluation of the number of qubit gates and circuit depth.} 

Table~\ref{tab-4} reports the reduction in two-qubit gate count and circuit depth achieved by TRAM relative to SABRE under the same compilation settings. 
On \textit{ibm-guadalupe},  relative to SABRE, TRAM reduces the total number of qubit gates by $10.83\%$, $15.38\%$, $7.48\%$, $22.63\%$, $2.54\%$, and $15.42\%$ for the six benchmark tasks, and shortens the circuit depth by $8.92\%$, $10.93\%$, $8.94\%$, $22.00\%$, $5.06\%$, and $18.00\%$, respectively. On \textit{ibm-brooklyn}, the corresponding gate-count reductions are $14.69\%$, $12.31\%$, $11.79\%$, $10.97\%$, $12.99\%$, $12.47\%$, $1.54\%$, and $9.80\%$, while the circuit depth decreases by $10.09\%$, $7.19\%$, $10.30\%$, $8.70\%$, $9.56\%$, $9.28\%$, $24.60\%$, and $18.39\%$, respectively. Overall, relative to SABRE, TRAM achieves an average reduction of $11.
49\%$ in gate count and $12.28\%$ in circuit depth on these two backends.

All experiments are compiled with optimization level~0 to isolate the effect of SWAP insertion from other compiler-level optimizations. Under this setting, any reduction in circuit depth or gate count directly reflects TRAM’s ability to minimize additional SWAP operations. Decoherence errors accumulate with circuit duration, and TRAM’s advantage lies in integrating these time-dependent errors into its initial qubit assignment. 
By constructing a time-weighted heatmap matrix and optimizing a global cost function that balances coupling fidelity and decoherence cost, TRAM identifies initial mappings that minimize SWAP usage while constraining circuit depth growth.

In Tab.~\ref{tab-3}, the absence of reduction achieved by TRAM in simulators can be attributed to two practical factors. First, for circuits with few qubits and shallow depth, the search space of valid mappings is small, and SABRE’s iterative heuristics can already find near-optimal solutions in terms of SWAP count. In contrast, TRAM’s objective emphasizes global fidelity rather than purely minimizing SWAP operations, which naturally limits further gate-count reduction in such small-scale circuits. Second, for shallow circuits, the impact of decoherence noise is weaker than depolarization noise. The THIM and T-SWAP components of TRAM jointly balance both error sources, compiling the circuit toward maximum overall fidelity even when gate count or depth remains unchanged. Consequently, TRAM still achieves fidelity improvement despite comparable structural complexity.

On larger backends such as $\textit{ibm-guadalupe}$ and $\textit{ibm-brooklyn}$, TRAM demonstrates clear advantages when compiling deep or highly entangled circuits. As circuit depth increases, decoherence effects dominate. While SABRE’s SWAP-reduction heuristics are  efficient, they lack explicit modeling of temporal noise accumulation. TRAM’s time-weighted routing approach addresses this gap: by assigning dynamic weights to two-qubit gates based on their execution order, it explicitly correlates the routing cost with decoherence magnitude. This time-aware cost model, combined with CQTP-based partitioning and global optimization, results in both shallower compiled circuits and higher effective fidelity across diverse workloads.

\medskip
\noindent\textbf{(c) Scalability and overall performance.} 

TRAM fully accounts for the impact of decoherence noise throughout the compilation process. It reduces SWAP insertion by selecting decoherence-aware initial mappings and further balances depolarization and decoherence effects during SWAP gate scheduling. 
Across all benchmark tests, TRAM achieves an average reduction of $11.49\%$ in the total number of qubit gates and $12.28\%$ in circuit depth relative to SABRE.

In addition to its quantitative gains, TRAM demonstrates strong scalability and robustness across different hardware configurations. It relies solely on the calibration data of the target quantum chip to construct effective qubit partitions, and experimental results confirm that these partitions remain stable across repeated calibrations. TRAM performs consistently well regardless of chip size or noise characteristics: 
on small-scale devices, it achieves higher circuit fidelity with compilation overhead comparable to SABRE, 
while on large-scale architectures, the community-detection-based CQTP efficiently handles complex coupling graphs. The hierarchical partitioning strategy of CQTP further enables parallel execution of multiple quantum programs on a single chip, enhancing hardware utilization and throughput. The integration of THIM and T-SWAP ensures that decoherence effects are explicitly modeled within both the partitioning and routing stages. For shallow circuits with limited entanglement, this design maintains fidelity parity with SABRE, whereas for deeper circuits containing extensive two-qubit interactions, 
it significantly suppresses decoherence-induced errors, reducing both circuit depth and total gate count. 

Overall, these results confirm that TRAM not only improves circuit fidelity but also provides a scalable, noise-adaptive compilation framework suitable for quantum devices of varying sizes and noise profiles.

	\section{Conclusion}\label{sec-conclusion}

\begin{figure*}[ht] 
  \centering  
  \subfigure[Comparison of fidelity between SABRE and TRAM.]{
   \includegraphics[width=1\textwidth]{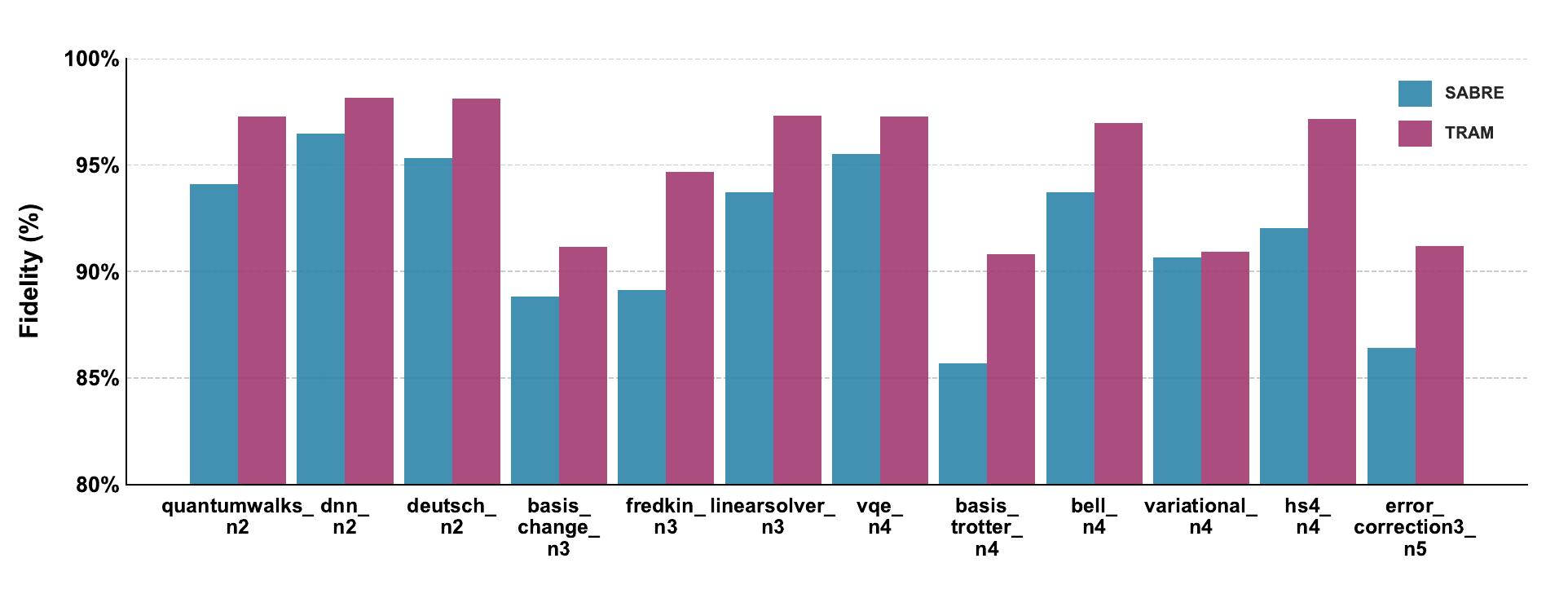}
  }
  \hfill 
  \subfigure[Comparison of the number of qubit gates between SABRE and TRAM.]{
   \includegraphics[width=1\textwidth]{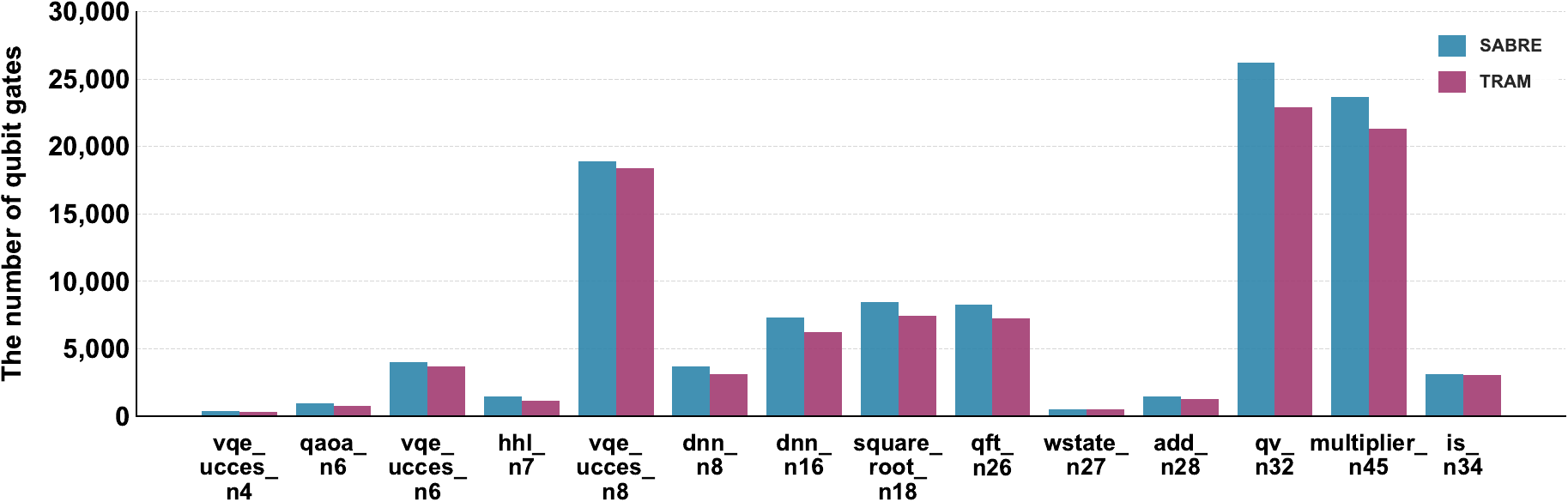}
  }
  \hfill  
  \subfigure[Comparison of circuit depth between SABRE and TRAM.]{
   \includegraphics[width=1\textwidth]{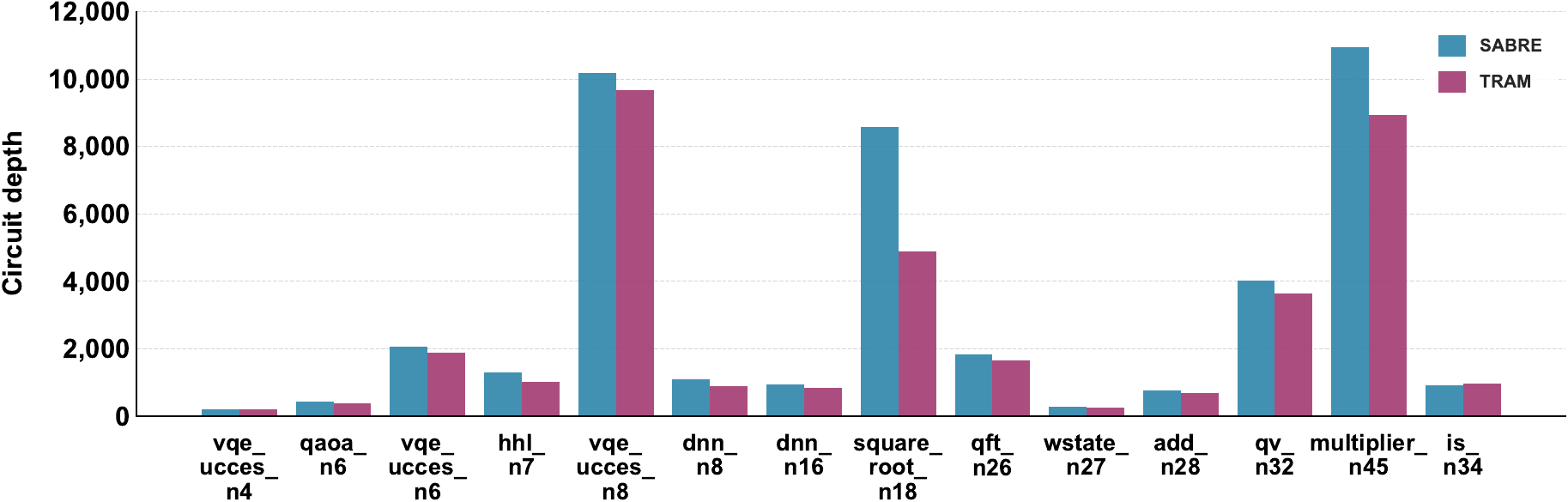}
  }
  \caption{\label{fig-11} Comparison of overall performance between SABRE and TRAM.
  (a) Fidelity comparison between SABRE and TRAM under identical benchmark tests, with quantitative results summarized in Tab.~\ref{tab-3}.
  (b) Qubit gate count comparison between SABRE and TRAM under the same benchmarks, with corresponding data presented in Tab.~\ref{tab-4}.
  (c) Circuit depth comparison between SABRE and TRAM under the same benchmarks, with detailed results provided in Tab.~\ref{tab-4}.
  }
\end{figure*}

To sum up, addressing the hardware constraints faced by quantum algorithms in NISQ devices and the issue that traditional qubit mapping ignores decoherence noise, this paper proposes the TRAM based on heuristic strategies. By integrating real quantum hardware data, this mechanism constructs highly correlated and noise-resistant partitions, generates an optimized initial mapping layout, and dynamically selects SWAP gate insertion positions based on decoherence noise, effectively making up for the shortcomings of traditional algorithms. Benchmark tests on the Qiskit framework simulator show that, compared with the existing SABRE algorithm, the TRAM algorithm has achieved steady improvement in fidelity, the number of qubit gates, and quantum circuit depth, as shown in Fig. \ref{fig-11}. This significantly optimizes the execution efficiency and reliability of quantum programs. The research results provide new ideas and effective approaches for overcoming the physical limitations of NISQ devices, improving the practical application performance of quantum algorithms, and developing execution tools for compilation. Future research will further explore how to combine real-time noise monitoring with adaptive adjustment strategies in more complex quantum hardware environments to continuously optimize the qubit mapping mechanism and promote the practicalization of quantum computing technology.

\section*{Acknowledgement}
This work was supported by the National Natural Science
Foundation of China (Grant No. 12271394).
This work was also supported by resources provided by the Pawsey Supercomputing Research Centre with funding from the Australian Government and the Government of Western Australia. It was carried out within the Pawsey Supercomputing Research Centre’s Quantum Supercomputing Innovation Hub, made possible by a grant from the Australian Government through the National Collaborative Research Infrastructure Strategy (NCRIS). Computational resources were provided by the Pawsey Supercomputing Research Centre’s Setonix Supercomputer (\href{https://doi.org/10.48569/18sb-8s43}{https://doi.org/10.48569/18sb-8s43}), with funding from the Australian Government and the Government of Western Australia.

\section*{DATA AVAILABILITY}
The data are not publicly available. The data are available from the authors upon reasonable request.

	\newpage
	\clearpage

	\newpage
	\clearpage
	
	\bibliography{SI}
	
\end{document}